\documentclass{aa}
\usepackage{graphicx}
\usepackage{txfonts}
\usepackage{verbatim}
\usepackage{amssymb,amsmath,amsfonts}

\usepackage{hyperref}
\hypersetup{
    colorlinks=true,
    citecolor=blue,
    linkcolor=blue,
    urlcolor=blue,
}

\usepackage{color}
\definecolor{byzantine}{rgb}{0.74, 0.2, 0.64}

\DeclareUnicodeCharacter{2212}{-}

\usepackage{natbib}
\bibpunct{(}{)}{;}{a}{}{,} 
 
\begin{document} 

   \title{TASTE V. A new ground-based investigation \\ of orbital decay in the ultra-hot Jupiter WASP-12b \thanks{Photometric data and dditonal tables are available in electronic form at CDS via anonymous ftp to
   cdsarc.u-strasbg.fr (130.79.128.5) or via http://cdsweb.u-strasbg.fr/cgi-bin/qcat?J/A+A/.}}

   \author{
   P.~Leonardi\thanks{E-mail: pietro.leonardi.1@studenti.unipd.it}
   \thanks{Email: pietro.leonardi@unitn.it} \inst{1,2} $^{\href{https://orcid.org/0000-0001-6026-9202}{\includegraphics[scale=0.5]{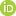}}}$ \and
   V.~Nascimbeni\thanks{E-mail: valerio.nascimbeni@inaf.it} \inst{3,2} $^{\href{https://orcid.org/0000-0001-9770-1214}{\includegraphics[scale=0.5]{FIGURES/orcid.jpg}}}$ \and
   V.~Granata \inst{4,3} $^{\href{https://orcid.org/0000-0002-1425-4541}{\includegraphics[scale=0.5]{FIGURES/orcid.jpg}}}$ \and
   L.~Malavolta \inst{2,3} $^{\href{https://orcid.org/0000-0002-6492-2085}{\includegraphics[scale=0.5]{FIGURES/orcid.jpg}}}$ \and
   L.~Borsato \inst{3}$^{\href{https://orcid.org/ 0000-0003-0066-9268}{\includegraphics[scale=0.5]{FIGURES/orcid.jpg}}}$  \and
   K.~Biazzo \inst{5} $^{\href{https://orcid.org/0000-0002-1892-2180}{\includegraphics[scale=0.5]{FIGURES/orcid.jpg}}}$ \and
   A. F.~Lanza \inst{6} $^{\href{https://orcid.org/0000-0001-5928-7251}{\includegraphics[scale=0.5]{FIGURES/orcid.jpg}}}$ \and
   S.~Desidera \inst{3} $^{\href{https://orcid.org/0000-0001-8613-2589}{\includegraphics[scale=0.5]{FIGURES/orcid.jpg}}}$ \and
   G.~Piotto\inst{2,3,4} $^{\href{https://orcid.org/0000-0002-9937-6387}{\includegraphics[scale=0.5]{FIGURES/orcid.jpg}}}$ \and
   D.~Nardiello \inst{2,3} $^{\href{https://orcid.org/0000-0003-1149-3659}{\includegraphics[scale=0.5]{FIGURES/orcid.jpg}}}$ \and
   M.~Damasso \inst{7} $^{\href{https://orcid.org/0000-0001-9984-4278}{\includegraphics[scale=0.5]{FIGURES/orcid.jpg}}}$ \and
   A.~Cunial \inst{2}\and
   L.~R.~Bedin \inst{3} $^{\href{https://orcid.org/0000-0003-4080-6466}{\includegraphics[scale=0.5]{FIGURES/orcid.jpg}}}$
          }

   \institute{Dipartimento di Fisica,  Università di Trento, Via Sommarive 14, 38123 Povo \and Dipartimento di Fisica e Astronomia, Universit\`a degli Studi di Padova, Vicolo dell'Osservatorio 3, 35122 Padova, Italy \and INAF -- Osservatorio Astronomico di Padova, Vicolo dell'Osservatorio 5, 35122 Padova, Italy \and Centro di Ateneo di Studi e Attività Spaziali “Giuseppe Colombo” (CISAS), Università degli Studi di Padova, Via Venezia 15, 35131 Padova, Italy \and INAF – Osservatorio Astronomico di Roma, Via Frascati 33, 00040 Monte Porzio Catone, Italy \and  INAF -- Osservatorio Astrofisico di Catania, via S. Sofia 78, 95123 Catania, Italy \and INAF - Osservatorio Astrofisico di Torino, Via Osservatorio 20, I-10025 Pino Torinese, Italy  }

   \date{Compiled: \today, Accepted: 16 February 2024}

  \abstract{The discovery of the first transiting hot Jupiters (HJs; giant planets on orbital periods shorter than $P\sim10$~days) was announced more than twenty years ago. As both ground- and space-based follow-up observations are piling up, we are approaching the temporal baseline required to detect secular variations in their orbital parameters. In particular, several recent studies focused on constraining the efficiency of the tidal decay mechanism to better understand the evolutionary time scales of HJ migration and engulfment. This can be achieved by measuring a monotonic decrease of orbital period $\mathrm{d}P/\mathrm{d}t<0$ due to mechanical energy being dissipated by tidal friction. WASP-12b was the first HJ for which a tidal decay scenario appeared convincing, even though alternative explanations have been hypothesized. Here we present a new analysis based on 28 unpublished high-precision transit light curves gathered over a twelve-year baseline and combined with all the available archival data, and an updated set of stellar parameters from HARPS-N high-resolution spectra, which are consistent with a main sequence scenario, close to the hydrogen exhaustion in the core. Our values of $\mathrm{d}P/\mathrm{d}t$ = $-30.72 \pm 2.67$ and  $Q_{\ast}^{'}$ = $(2.13 \pm 0.18) \times 10^{5}$ are statistically consistent with previous studies, and indicate that WASP-12 is undergoing fast tidal dissipation. We additionally report the presence of an excess scatter in the timing data and discuss its possible origin.}


   \maketitle
%
\section{Introduction}

The discovery of Jupiter-sized extrasolar planets orbiting their host stars at unexpectedly close distances (``hot Jupiters"; HJ) posed several new questions about planetary formation and planet migration theories.
Over the last two decades, astronomers have taken important steps to understand how HJs originated. As of now, the main process suggested to be responsible for such short orbits is migration from their formation site further out (\citealt{dawson}; \citealt{fortney}). The two most accredited transport mechanisms are \emph{disk-driven migration}, due to friction with the gaseous disk around the young star (\citealt{lin}; \citealt{nelson}), and \emph{high-eccentricity} migration (HEM),  through planet-planet scattering \citep{rasio} or secular interactions (\citealt{fabrycky}; \citealt{wu}).
If we consider the current occurrence estimates of one planet every two stars in the Milky Way (\citealt{howard}; \citealt{dressing}; \citealt{batalha}; \citealt{silburt}), it is clear that only a tiny fraction of the total exoplanet population has been detected and studied.
The many planetary systems identified so far range in size from tiny rocky planets to massive gas giants, and showing us that diversity is one of the most important aspects of exoplanet demographics. In recent years a new subclass of HJs has emerged, the so-called ``ultra-hot Jupiters'' (UHJs), having day-side temperatures higher than 2200 K \citep{parementier} and orbital periods between $\lesssim 1$ day up to $\sim 2$ days. Given their extreme properties, they are valuable test cases for probing atmospheric chemistry \citep{monika}, understanding planetary mass loss, and constraining planetary formation and evolution models.

WASP-12b, the subject of this study, is one of such UHJs ($T_\mathrm{eq}\simeq 2600$ K). Discovered in 2009 by \citet{hebb}, as part of the Wide-Angle Search for Planets (WASP) project, WASP-12b is known to be inflated ($R_{p}=1.825 \pm 0.091  \ R_\mathrm{jup}$; $M_{p}=1.39 \pm 0.12$~$M_\mathrm{jup}$) and orbiting a 
an F-type star ($T_\mathrm{eff}=6250$ K) with a small projected rotational velocity  ($v\sin i=2.2 \pm 1.5 $ km/s, \citealt{albrecht})  in $P\simeq1.09$ days. 
The peculiar architecture of WASP-12b has inspired several research studies on its atmosphere and planet-star interactions.
Strong absorption lines of metals in the near-UV spectra were revealed by the Cosmic Origins Spectrograph (COS) on the Hubble Space Telescope (\citealt{fossati}; \citealt{haswell}; \citealt{nichols}). Optical transit measurements unveiled that the planet is inflated, and infrared phase curves showed that it is filling $\sim 84\%$ of its Roche lobe resulting in high mass-loss rates of metals and heavy elements  (\citealt{lai}; \citealt{li}; \citealt{bell2019}; \citealt{Antonetti&Goodman_2022}). An escaping exosphere was suggested as the cause of these phenomena (\citealt{li}; \citealt{fossati}; \citealt{haswell}). According to hydrodynamic simulations by \cite{debrecht}, a circumstellar disk is forming due to the expanding exospheric gas that overfills the Roche lobe and escapes via the Lagrangian L1 and L2 points. Finally, a new atmospheric model also including the formation of $\textrm{H}^{-}$ through dissociation of $\textrm{H}_{2}\textrm{O}$ and $\textrm{H}_{2}$ has been proposed by \citet{Himes2022} to reconcile the previous findings.

Another notable feature of WASP-12b is its changing orbital period.
A single planet on an unperturbed Keplerian orbit should transit at regular intervals, i.e. having a perfectly constant orbital period. If the transits are no longer strictly periodic, among the possible causes for this perturbation there are the presence of additional bodies in the system,  tidal forces and relativistic effects (e.g., \citealt{agol1}; \citealt{holman}; \citealt{antoniciello}). On close-in planets, tidal effects can cause orbital circularization and a change in orbital period (\citealt{rasio}; \citealt{levrard}). Since the early days of exoplanet research (\citealt{rasio}; \citealt{sasselov}), it has already been suggested that short-period planets could be unstable due to tidal dissipation. Detection of orbital decay is now within reach because many transiting exoplanets, predominantly HJs, have been monitored for decades, using both space- and ground-based surveys, as part of long-term transit timing observation campaigns.  

The first claim of short-term period variations in WASP-12b was published by \citet{macie2011} and interpreted as due to the dynamical perturbation from an unseen planetary companion (so-called Transit Time Variations; TTV). This detection prompted other observational campaigns to confirm and assess the cause of this signal.
\cite{mac2016} published additional data and discovered a quadratic trend in the $O-C$ diagram of the mid-transit timing compatible with a uniformly decreasing orbital period. Following studies validated this finding (\citealt{patra2017}; \citealt{collins}; \citealt{mac2018}; \citealt{baluev}), which might be equally explained by 1) an orbital shrinkage related to tidal decay or 2) the Rømer effect from an unseen companion or 3) part of a long-term ($\simeq 14$~yrs) oscillation generated by apsidal orbital precession. The latter hypothesis is plausible if the eccentricity is $e \simeq 0.002$. However, due to the expected fast rate of tidal circularization, it is unclear how to retain such an eccentricity \citep{weinberg}. \cite{yee} presented compelling evidence against apsidal precession and the Rømer effect as major contributors. 
The authors demonstrated how the $O-C$ values of the occultations have a decreasing pattern compatible with that of the $O-C$ values of the transits, while one should expect the mid transits and occultations to exhibit an opposite trend when the apparent orbital period variation is due to the apsidal precession.
Furthermore, their study revealed no significant acceleration of the barycenter of the system along the line of sight using radial velocity observations, disproving the Roemer effect as the primary cause of the observed decreasing $O-C$ trend.
This supported the orbital decay scenario, further confirmed by the two latest studies on this system from the TESS space telescope primary and extended missions: \citet{turner} and \citet{wong}. The latter research derived a value for the period change rate $dP/dt =-29.81 \pm 0.94$~ms/yr by including transit and secondary eclipse light curves.

In this work we present 28 unpublished, high-precision, ground-based light curves of WASP-12b.
Additionally, we analyze high-resolution optical spectra from the High Accuracy Radial velocity Planet Searcher in the Northern hemisphere (HARPS-N, \citealt{cosentino2012}; \citeyear{cosentino2014}) at the Telescopio Nazionale Galileo (TNG) in La Palma, to derive the stellar parameters of the host star and improve the calculation of the tidal quality factor. 
In Section \ref{section:obs_data_reduction} of this paper, we present the new photometric and spectroscopic observations and our data reduction procedures. In section \ref{section:data_analysis}, we describe the data analysis.
The results of the timing analysis aimed at computing the orbital period change rate and constraining the stellar tidal quality factor $Q_{\ast}^{'}$ are reported in Section \ref{section:discussion}. Finally, in Section \ref{section:summary} we present a summary of our conclusions. \\

\section{Observations and Data Reduction} 
\label{section:obs_data_reduction}

\subsection{Photometry}
\label{photometry}

The unpublished photometric observations we present were gathered by the TASTE project, a long-term observing program aimed at monitoring transiting planets and creating a library of high-precision light curves to exploit the TTV (Time Transit Variation) technique (\citealt{nascim}; \citealt{granata_2014}). The project is mostly based on the Asiago Astrophysical Observatory located at Mount Ekar (elevation: 1366~m) in northern Italy, plus other medium-class telescopes around the globe. The Asiago facilities include the 1.82m Copernico telescope, a Cassegrain telescope equipped with the Asiago Faint Object Spectrograph and Camera (AFOSC) with a $8.8 \times 8.8$ arcmin$^{2}$ field of view, which acquired all but one of our light curves. The AFOSC detector is a back-illuminated $2048 \times 2048$ E2V pixels CCD. The remaining light curve was obtained at the 67/92 cm Schmidt telescope at the same Observatory. Our observing strategy is based on defocusing the images and applying differential photometry techniques to minimize the impact of systematic errors from instrumental and telluric sources.
For all observations, we adopted for consistency the same set of 2 reference stars (TYC 1891-38-1 and TIC 86396443), always imaged over the years in the field of view of the telescope. There are no known variability indicators or higher-than-expected photometric scatter in any of them. A complete description of the TASTE observing strategy and data reduction software is given by \cite{nascim,nascimbeni2013}.

Our 28 transit light curves were collected over 12 years (2010-2022); a detailed observing log is reported in Table \ref{table:log_nights}, summarizing the dates and other relevant information. All the images were acquired either with a standard Cousins $R_C$ filter ($\lambda_\mathrm{eff} = 672.4$~nm) or a standard SDSS $r'$ ($\lambda_\mathrm{eff} = 620.4$~nm), both chosen to maximize the atmospheric extinction and limb darkening effects. The advantage of observing in the $R_C$ or $r'$ passbands is that the impact of the limb-darkening on the transit profile is reduced with respect to the V passband, thus the timing of the ingress and egress is improved. Moreover, extinction in the Earth's atmosphere is also reduced by observing in those passbands.  We applied windowing and $4 \times 4$ binning to increase the duty cycle of the series as much as possible, while keeping our reference stars within the imaged field. Stars were intentionally defocused to an approximate radius of about 3~arcsec, equivalent to $\sim$ 12 physical pixels, to avoid saturation and to minimize systematic errors due to pixel inhomogeneities and imperfect flat-field correction. The exposure time ranged between 2 to 6~s, according to the weather conditions and the amount of defocus applied. For each observation, an out-of-transit part was secured about one hour before ingress and one hour after egress for normalization and detrending purposes. Bias and flat-field frames were collected according to the standard data reduction techniques. 
We carried out a preliminary selection of our frames, checking for any quality issues  (e.~g., pixels exceeding the saturation level, strong cosmic-ray hits, etc.). After identifying the problematic frames, we removed the corresponding data points from each light curve.
In six of the twenty-eight observations, clouds, high humidity, veils, and other weather conditions prevented a full complete transit from being observed.  We did not exclude from our analysis the partial transits
when at least the ingress or the egress were well sampled, since most of the timing information lies at those phases where the time derivative of the flux is larger.

\begin{table*}\centering\small\renewcommand{\arraystretch}{1.2}
    \caption{Observing log of our unpublished transit light curves of WASP-12b. }
    \begin{tabular}{c c c c c c c | l}
    \hline\hline
    ID  &  Evening date  & Telescope & $N$ & Filter & Exptime [s] & RMS [mmag] &  Phase coverage \\
    \hline
    \texttt{1} & 2010-12-11 & 1.82-m & 2996  & $R_{C}$ & 4-5  & 3.46 & Full transit  \\
    \texttt{2} & 2010-12-12 & 1.82-m  & 2366  & $R_{C}$ & 4-5-7 & 4.55 &  Full transit \\
    \texttt{3} & 2011-02-06 &  Schmidt & 1223 & $R_{C}$ & 10 & 4.11 & Full transit \\
    \texttt{4} & 2011-11-22 & 1.82-m   & 494    & $R_{C}$ & 10 & 1.96 & Full transit \\
    \texttt{5} & 2011-11-23 & 1.82-m     & 2393  & $R_{C}$ & 5  & 1.81& Full transit \\
    \texttt{6} & 2011-12-17 & 1.82-m   & 1881   &    $R_{C}$ & 4-7   & 3.20 &  Full transit \\
    \texttt{7} & 2011-12-27 & 1.82-m  & 3077   &    $R_{C}$ & 4   & 2.95 & Full transit  \\
    \texttt{8} & 2011-12-28 & 1.82-m   & 2456   &   $R_{C}$ & 5     & 2.32 & Full transit \\
    \texttt{9} & 2012-02-24 & 1.82-m   &    2449  & $R_{C}$ & 5   & 2.84  & Full transit\\
    \texttt{10}& 2012-10-21 & 1.82-m  & 1881 & $R_{C}$ &5 & 2.55 &  Partial (egress only) \\
    \texttt{11}& 2012-12-19 & 1.82-m & 2268 & $R_{C}$ & 5& 1.75  & Full transit \\ 
    \texttt{12}& 2013-01-11 & 1.82-m &	2287 & $R_{C}$ &4-5-7 & 2.48  & Full transit \\
    \texttt{13}& 2013-02-04	& 1.82-m &	2509 & $R_{C}$ &5 &  2.50  & Full transit \\
    \texttt{14}& 2013-12-02	& 1.82-m &	2784 & $R_{C}$ & 4& 2.77 &  Partial (only ingress) \\
    \texttt{15}& 2014-03-05	& 1.82-m &	1945 & $R_{C}$ & 4& 4.13 &  Partial (only ingress)\\
    \texttt{16}& 2015-12-11	& 1.82-m &	2680 & SDSS $r'$ & 3& 2.81 &  Partial (only egress) \\
    \texttt{17}& 2015-12-12	& 1.82-m &	3204 & SDSS $r'$ & 2& 2.68 & Full transit \\
    \texttt{18}& 2017-01-20	& 1.82-m &	2646 & SDSS $r'$ & 3& 3.01 & Full transit \\
    \texttt{19}& 2017-12-20	& 1.82-m &	2084 & SDSS $r'$ & 5& 2.56  & Full transit\\
    \texttt{20}& 2018-01-12	& 1.82-m &	2437 & SDSS $r'$ & 4& 2.90 & Full transit  \\
    \texttt{21}& 2019-12-29	& 1.82-m &	2376 & SDSS $r'$ &5 & 1.70 & Full transit \\
    \texttt{22}& 2020-01-22	& 1.82-m & 	3080 & SDSS $r'$ &4 & 2.29 & Full transit \\
    \texttt{23}& 2020-11-06	& 1.82-m &	4045 & SDSS $r'$ & 2&  3.87 & Full transit \\
    \texttt{24}& 2020-11-17	& 1.82-m &	2942 & SDSS $r'$ & 4& 2.69& Full transit \\
    \texttt{25}& 2021-10-29  & 1.82-m & 2419 & SDSS $r'$ & 5& 3.23 & Full transit \\ 
    \texttt{26}& 2021-12-28 & 1.82-m & 1805 & SDSS $r'$ & 4& 3.51 & Partial (egress only) \\
    \texttt{27}& 2022-01-08 & 1.82-m & 1350 & SDSS $r'$ & 4-5& 2.56 & Partial (ingress only) \\
    \texttt{28}& 2022-03-08 & 1.82-m & 2845 & SDSS $r'$ & 4& 2.73 & Full transit \\
    \hline
    \end{tabular}
    \label{table:log_nights}
    \tablefoot{The columns show: a numerical identifier, the evening date of the observation, the telescope used, the number $N$ of valid scientific frames acquired during the observation, the photometric filter, the average exposure time in seconds, the RMS of the residuals and the approximate phase coverage.}
\end{table*}

\subsection{Spectroscopy}

WASP-12 was observed with HARPS-N at TNG at a spectral resolution of $110\,000$, in the framework of the GAPS project \citep{covino2013}.
A total of 51 HARPS-N spectra, taken from November 2012 to January 2018, were used in this study.
An analysis including the first 15 spectra was published by \cite{bonomo} together with the result of the other 44 stars observed in that study. 
Additionally, we used eight spectra from the same program, which are still unpublished.
The main purposes of these first 23 observations were to refine the planetary parameters and to search for additional companions through high-precision RV monitoring (individual RV errors ranging from 2 to 10 m/s).
The remaining 28 spectra were obtained during planetary transits and adjacent off-transit windows on the nights 2017-12-23 and 2018-01-14 as part of the GAPS program to study the planetary atmospheres of hot Jupiters \citep{guilluy}.
The individual spectra were reduced with the standard Data Reduction Software version 3.8.
A coadded spectrum was then created using all the spectra, achieving a mean S/N of $\sim 220$ at around $6000$\,\AA.
The HARPS-N spectra described above were exploited in this paper to characterise the host star (see Sect.\,\ref{section: stellaparameters}).
For WASP-12, there are no indications of additional companions, and the updated RV analysis yielded a RV semi-amplitude of $219.90 \pm 2.2$ m/s, corresponding to a planetary mass of $1.39 \pm 0.12$ $M\mathrm{jup}$, with no significant eccentricity ($e<$0.02).

\section{Data analysis}
\label{section:data_analysis}

\subsection{Light curve extraction}

After the images were corrected for bias and flat field, we performed differential aperture photometry and extracted the light curves by employing the  \texttt{STARSKY} photometric pipeline, originally designed for the TASTE project (\citealt{nascim}; \citealt{nascimbeni2013}), and based entirely on an empirical approach. This pipeline performs aperture photometry on the target and a set of reference stars, and combine their fluxes to get high-precision differential light curves, where instrumental or telluric systematic errors are mitigated. 
The error bars on each data point were calculated with the analytic formulae quoted by \citet{nascim} (Eq.~1 and 2) and through standard statistical  propagation.

The aperture radius was set to include most of the flux of the target and the reference stars, and to minimize the off-transit scatter of each light curve. The optimal radius fell in the range of 9-23 pixels.
All observations utilized the same set of two reference stars as indicated in Sect.~\ref{photometry} for consistency. Before the extraction, all the light curves were detrended with a linear function of time at the extraction stage, as part of the  \texttt{STARSKY} standard processing (\citealt{nascim}; \citealt{nascimbeni2013}).

In order to represent all our time stamps to a single, accurate time standard, we translated them to the mid-exposure instant and converted them to the BJD-TDB standard, following the prescription by \citet{eastman}.
It is worth mentioning the light curve ID=\texttt{23} is the only one in our set showing an unexpected feature, a 5-mmag deep, 6-minute long dip right after the end of ingress (see Fig.~\ref{figure:thirdlc}). Although the overall RMS of this light curve is higher than average  due to a high background level (bright moon at $\sim20^\circ$), there is no clear correlation between the feature and any of our diagnostic parameters, including guiding drifts, stellar FWHM, transparency, background, etc. The dip is independent of the choice of reference stars, and is even visible in the absolute light curve. We conclude that the feature is probably genuine, and due to the crossing of an active region during the transit. We will come back to this explanation in Section~\ref{section:scatter_data}.

\subsection{Light curve modeling}
\label{section_fitting}

We analyzed the whole set of twenty-eight transit light curves simultaneously using the package PyORBIT\footnote{\url{https://github.com/LucaMalavolta/PyORBIT}} (\citealt{malavolta2016}; \citeyear{malavolta2018}) developed for modelling planetary transits and radial velocities.
In the fitting procedure, the central time of transit $T_{0}$ was left free to vary  for every transit. As each data set includes a single transit, a uniform prior on each central time of transit T$_{0}$ is automatically generated by taking as boundaries the first and last epoch of the corresponding data set. In those cases where only a partial light curve has been observed and the $T_{0}$ falls outside the observed window (namely, ID\# \texttt{10}, \texttt{14}, \texttt{15}, \texttt{16}, \texttt{27}), the uniform prior was set manually via  visual inspection.

We imposed a Gaussian prior on the orbital period ($P$) and the stellar radius, based on the values from \citet{bonomo}.
We adopted a quadratic law to model WASP-12's limb darkening (LD) effect, employing the limb darkening parameterization ($q_{1}$, $q_{2}$) introduced by \cite{kipping2013}.
The limb darkening coefficients are associated with two sets of Gaussian priors, one for each filter, centred on the prediction of the PHOENIX atmospheric models \citep{husser} with an added conservative uncertainty of 0.10.

We assumed a circular orbit. The effect of non-zero eccentricity on the light curve implies a difference in the ingress and egress times. This difference is $10^{-2} \times e$ for a close-in planet \citep{winn_2010}. Since our eccentricity is < 0.02 (See Sect. \ref{section: stellaparameters}) the effect on the light curve is negligible. Additionally, for a short period planet orbiting a star as old as WASP-12 (See Table \ref{tabble:stellar_parameters}), we can consider that the orbit is circularized (\citealt[][Fig. 4]{nagasawa}).

For each data set, we included a jitter term to account for possible underestimation of measurement errors and short-term stellar activity white noise.
We additionally used a quadratic baseline (three extra free parameters for each light curve) for transits ID\# \texttt{3}, \texttt{4}, \texttt{5}, \texttt{14}, \texttt{15}, \texttt{21}, and \texttt{22}, i.e., for those where the corresponding BIC value indicated a significant improvement in fit as a result of the inclusion of the detrending baseline.
Finally, because no contaminants fall inside the photometric aperture adopted, the dilution factor is negligible, and is it not included in the fit.

The free parameters in the fit accounted for: three parameters for the transit shape, four for the LD parameters, 28 for the transit times, 28 for the jitter and 15 ($3 \times 15$) for the detrending parameters. This amounts to a total of 78 free parameters.

All the transit models were computed with the popular package \texttt{batman} \citep{kreidberg}, with an exposure time of 5 s and an oversampling factor of 1. 
The differential evolution algorithm PyDE\footnote{\url{https://github.com/hpparvi/PyDE}} was used to conduct global parameter optimization. The output parameters served as the starting point for the Bayesian analysis, which was carried out using the \texttt{emcee} \citep{foreman-mackey} package, a Markov Chain Monte Carlo algorithm. We ran an auto-correlation analysis on the chains: if the chains were longer than 100 times the estimated auto-correlation time and this estimate varied by less than 1\%, the chains were deemed converged. We cautiously set the burn-in value to a number greater than the previously stated convergence point, and we used a precautionary value of 1000 for the thinning factor. We ran the sampler for $75\,000$ steps, with 680 walkers and a burn-in cut of $25\,000$ steps.
Table~\ref{table:T0} shows all the 28 transit central times obtained from the PyORBIT fit. The resulting best-fit parameters and the priors are shown in Table \ref{table:orbital_parameters_wasp12b}. The light curves and the corresponding PyORBIT best-fitting models are plotted in Figures \ref{figure:firstlc}-\ref{figure:lastlc}.

\begin{table}[tb]
    \small\centering\renewcommand{\arraystretch}{1.2}
    \caption{New best-fit transit times of WASP-12b.}
    \begin{tabular}{c c r}
    \hline\hline
    ID & $T_0$ (BJD-TDB) & $\sigma_{T0}$ [days] \\
    \hline
    \texttt{1} & 2455542.553168    &     0.000189 \\
    \texttt{2} & 2455543.643326      &   0.000297\\
    \texttt{3} & 2455599.307161     &    0.000540\\
    \texttt{4} & 2455888.533480    &    0.000180\\
    \texttt{5} &  2455889.624786    &     0.000135\\
    \texttt{6} & 2455913.635896     &    0.000204\\
    \texttt{7} &  2455923.458691     &    0.000164\\
    \texttt{8} & 2455924.549735     &    0.000125\\
    \texttt{9} &2455982.395378    &    0.000160\\
    \texttt{10} & 2456222.508576      &    0.000210\\
    \texttt{11} & 2456281.444317      &    0.000103\\
    \texttt{12} & 2456304.364434      &    0.000133\\
    \texttt{13} & 2456328.375176       &   0.000143\\
    \texttt{14} & 2456629.607692      &   0.000266\\
    \texttt{15} &2456722.377750      &    0.000272\\
    \texttt{16} &2457368.498222        &   0.000225\\
    \texttt{17} &2457369.589776        &   0.000125\\
    \texttt{18} &2457774.507194       &    0.000156\\
    \texttt{19} &2458108.480490       &    0.000148\\
    \texttt{20} &2458131.399324      &     0.000171 \\
    \texttt{21} &2458847.370515     &     0.000186\\
    \texttt{22} &2458871.381793    &      0.000155\\
    \texttt{23} &2459160.606724     &     0.000164\\
    \texttt{24} &2459171.521679     &     0.000132\\
    \texttt{25} &2459517.500786      &    0.000194\\
    \texttt{26} &2459577.528473      &    0.000253\\
    \texttt{27} &2459588.443252      &    0.000175\\
    \texttt{28} &2459647.379007      &    0.000128 \\
    \hline
    \end{tabular}
    \tablefoot{The numerical IDs match those assigned on Table~\ref{table:log_nights}. The transit times are given in the BJD-TDB standard \citep{eastman}; the third column reports the associated 1-$\sigma$ error.}
    \label{table:T0}
\end{table}

\begin{table*}
    \centering\small\centering\renewcommand{\arraystretch}{1.3}
    \caption{WASP-12b orbital parameters derived from PyORBIT Markov Chain Monte-Carlo analysis.} 
    \begin{tabular}{l c c r}
    \hline\hline
    Parameter & Units &  Prior &  Posterior value \\
    \hline
    \textsc{Fitted Parameters} & & &\\
    Orbital Period ($P$) & [days] &  $\mathcal{U}[1.09142090, 0.00000020]$ & $1.0914210_{-0.00000020}^{+0.00000019}$ \\
    Impact Parameter ($b$) &-- & $\mathcal{U}[0, 1+ (R_{\star}/R_{p})/ 2]$ & ${0.424}_{-0.013}^{+0.012}$ \\
    Quadratic Limb Darkening term $c_{1}$ ($R_{C}$ filter) &-- & $\mathcal{U}[0.412 ,0.10]$ & $0.349_{-0.029}^{+0.029}$\\ 
    Quadratic Limb Darkening term $c_{2}$ ($R_{C}$ filter) & --& $\mathcal{U}[0.148 ,0.10]$ &  $0.167_{-0.049}^{+0.049}$\\ 
    Quadratic Limb Darkening term $c_{1}$ (SDSS.r filter) & --& $\mathcal{U}[0.428 ,0.10]$ & $0.369_{-0.031}^{+0.030}$ \\ 
    Quadratic Limb Darkening term $c_{2}$ (SDSS.r filter) & --&  $\mathcal{U}[0.15 ,0.10]$ & $0.109_{-0.051}^{+0.053}$\\ 
    \\
    \textsc{Derived Parameters} &  & & \\ 
    Semi-major axis ($a$) & [au] & --  & ${0.0234}_{-0.0010}^{+0.0010}$ \\ 
    Inclination ($i$) & [deg] &--  & ${81.80}_{-0.27}^{+0.29}$ \\ 
    Planet Radius ($R_{p}$) & [$R_{J}$]  & --  & ${1.965}_{-0.087}^{+0.088}$ \\ 
    Transit duration ($T_{14}$) & [days] & --& $0.12504 \pm 0.00018$ \\ 
    \hline
    \end{tabular}
    \label{table:orbital_parameters_wasp12b}
    \tablefoot{The listed best-fit values and uncertainties are the medians and 15.865th-84.135th percentiles of the posterior distributions, respectively.}
\end{table*}

\subsection{Stellar parameters}
\label{section: stellaparameters}
The coadded spectrum was analyzed as in \citet{biazzo2022} to derive the effective temperature $T_{\rm eff}$, the surface gravity $\log g$, the microturbulence velocity $\xi$, the iron abundance [Fe/H], and the rotational velocity $v\sin{i_{\star}}$ of WASP-12. For $T_{\rm eff}$, $\log g$, $\xi$, and [Fe/H] we applied a method based on equivalent widths (EWs) of 81 \ion{Fe}{i} and 11 \ion{Fe}{ii} lines taken from \cite{Biazzoetal2015} and \cite{biazzo2022} and the spectral analysis package MOOG (\citealt{Sneden1973}; version 2017). We then adopted the \cite{CastelliKurucz2003} grid of model atmospheres with solar-scaled chemical composition and new opacities. $T_{\rm eff}$ and $\xi$ were derived by imposing that the abundance of the \ion{Fe}{i} lines is not dependent on the line excitation potentials and the reduced equivalent widths (i.e. $EW$/$\lambda$), respectively, while $\log g$ was obtained by imposing the  ionization equilibrium condition between the abundances of \ion{Fe}{i} and \ion{Fe}{ii} lines.

The $v\sin{i_{\star}}$ was measured with the same MOOG code and applying the spectral synthesis technique of three regions around 5400\,, 6200\,and 6700\,\AA. We adopted the grid of model atmosphere mentioned above and, after fixing the macroturbulence velocity to the value of 5.5~km/s from the relationship by \cite{Doyleetal2014}, we find an upper limit for the stellar $v\sin{i_{\star}}$ of $1.9$~km/s.

From the same coadded spectrum, we also derived the abundance of the lithium line at $\sim$6707.8\,\AA\, $\log A$(Li), after measuring the Li EW ($39.5\pm 1.5$~$\AA$) and considering our stellar parameters previously derived together with the non-LTE corrections by \cite{Lindetal2009}. We therefore obtained $2.55\pm 0.05$~dex as NLTE lithium abundance. Lithium equivalent width and elemental abundance values for a star with effective temperature like WASP-12 are intermediate between those of clusters of 2 Gyr, such as NGC752, and $\sim$4 Gyr, such as M67 \citep{Sestitoetal2005, Jeffriesetal2023}. Final atmospheric stellar parameters, together with iron and lithium elemental abundances are listed in Table\,\ref{table:derived_stellar_parameters}, where uncertainties were computed as in \cite{biazzo2022}.

In order to  determine the mass, radius, density and age, we followed the same approach as described in Sect. 3.2 of \cite{lacedelli21}. 
We used the code \textsc{isochrones} \citep{morton}, with posterior sampling through the code {\textsc{MultiNest}} \citep{feroz08,feroz09,feroz19}.
We provide as input the photometry from the Two Micron All Sky Survey 2MASS, WISE, plus the parallax of the target from the Gaia eDR3 (See Table \ref{tabble:stellar_parameters} for the adopted stellar parameters). Analysis has been performed using two evolutionary stellar models, the MESA Isochrones \& Stellar Tracks (MIST) \citep{dotter16,choi,paxton} and the Dartmouth Stellar Evolution Database \citep{dotter08}.

We derived from the mean and standard deviation of all posterior samplings, $M_{\star}$ = $1.325^{+0.026}_{-0.018}$ $M_{\odot}$, $R_{\star}$ = $1.690_{-0.018}^{+0.019}$ $R_{\odot}$. The stellar density $\rho_{\star}$ = $0.27 \pm 0.010$ $\rho_{\odot}$ was derived directly from the posterior distributions of $M_{\star}$ and $R_{\star}$.  These stellar parameters, together with the age, are listed in Table \ref{table:derived_stellar_parameters}.

\begin{table}[tb]
\centering\centering\renewcommand{\arraystretch}{1.4}\small
    \caption{Stellar parameters of WASP-12 derived from the analysis of the HARPS-N spectra and from stellar evolutionary tracks.} 
    \begin{tabular}{l c}
    \hline\hline
    Parameter & Value\\
    \hline
    T$_\mathrm{eff}$ [K] & $6265 \pm 50$ \\
    $\log g$ [cgs] & $4.11 \pm 0.11$ \\
    $[\mathrm{Fe}/\mathrm{H}]$\,[dex] & $0.12 \pm 0.07$ \\
    $\xi$ [km s$^{-1}$] & $1.39 \pm 0.08$\\
    $v \sin i_{\star}$ [km s$^{-1}$] & $< 1.9$ \\
    $EW_{\rm Li}$ [m\AA] & $39.5 \pm 1.5$   \\
    $\log A$(Li) [dex] & $2.55 \pm 0.05$  \\
    $A_{V}$ [mag] & $0.321 \pm 0.096$ \\
    $M_{\star}$ [$M_{\odot}$] & $1.325^{+0.026}_{-0.018}$ \\ 
    $R_{\star}$ [$R_{\odot}$] & $1.690_{-0.018}^{+0.019}$ \\
    $\rho_{\star}$ [$\rho_{\odot}$] & $0.274 \pm 0.010$ \\
    Age [Gyr] & $3.05 \pm 0.32$ \\
    \hline
    \end{tabular}
    \label{table:derived_stellar_parameters}
\end{table}

\begin{table}[tb]
    \centering\small
    \caption{Astrophysical properties of WASP-12}
    \begin{tabular}{l   c   r}
    \hline\hline
      Parameter & Value & Reference  \\
      \hline
      \textit{Identifiers} & & \\
      2MASS &  J06303279+2940202 & 1\\ 
      Gaia DR3 & 3435282862461427072 & 2\\
      \\
      \textit{Astrometric Parameters} & & \\
      RA [J2000] & 06:27:21.28 &  2\\
      Dec [J2000] & 29:42:26.84 &  2\\
      Parallax [mas] & $2.4213 \pm 0.0166$ &  2\\
      $\mu_{\alpha}$ [mas $yr^{-1}$] &  $-1.518 \pm 0.019$ & 2\\
      $\mu_{\delta}$ [mas $yr^{-1}$] & $-6.760 \pm 0.015$ & 2 \\ 
      \\
      \textit{Photometric parameters} & & \\
      B[mag] & $12.14 \pm 0.21$ & 3 \\
      V[mag] & $11.57 \pm 0.16$ & 3 \\
      J[mag] & $10.477 \pm 0.021$ & 1 \\
      H[mag] & $10.228 \pm 0.022$ & 1 \\
      K[mag] & $10.188 \pm 0.020$ & 1 \\
      W1[mag] & $10.111 \pm 0.022$ & 4 \\
      W2[mag] & $10.109 \pm 0.021$ & 4 \\ 
      W3[mag] & $10.121 \pm 0.073$ & 4 \\
    \hline     
    \end{tabular}
    \tablebib{\\
    (1) Two Micron All Sky Survey (2MASS; \citealt{cutri_2003}; \citealt{skrutskie_2006}); \\ (2) Gaia eDR3 \citep{gaia_eDR3_2020} \\ (3) Tycho-2 Catalogue \citep{tycho-2} \\ (4) Wide-field Infrared Survey Explorer (WISE; \citealt{wright_2010}) }

    \label{tabble:stellar_parameters}
\end{table}

\subsection{Timing Analysis}
\label{timing_analysis}
\begin{figure*}[tb]
\sidecaption
    \includegraphics[width=1.4\columnwidth]{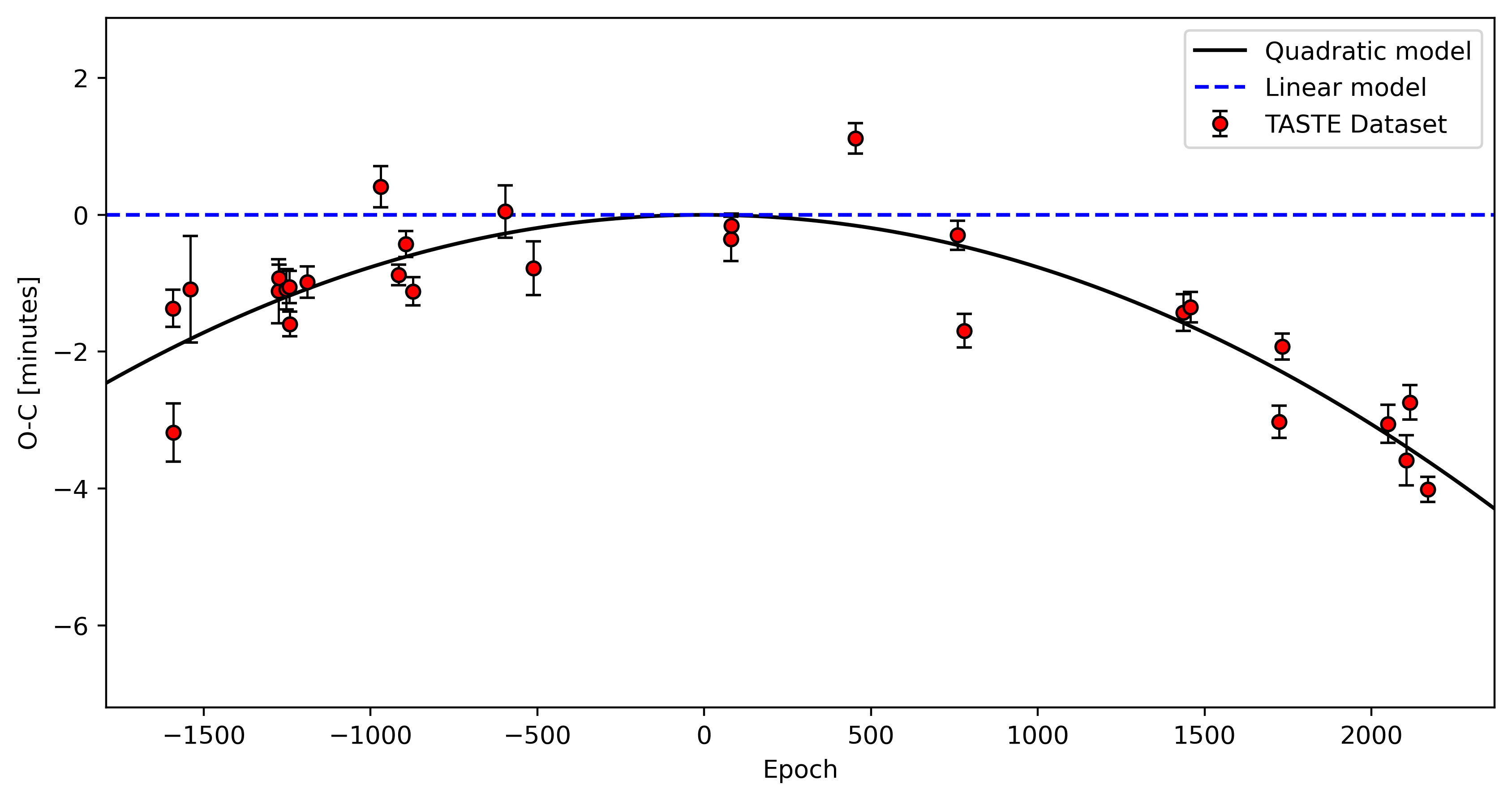}
    \caption{Observed minus calculated ($O-C$) diagram of WASP-12b for the unpublished transits collected at Asiago (Tables~\ref{table:log_nights} and \ref{table:T0}). The reference ephemeris here is the linear part of Eq.~\ref{eq:asiagofit}, while the additional quadratic term (based on d$P$/d$E$) is plotted as a black solid line. An excess scatter in the data points with respect to the quadratic model can be seen in the figure. This is addressed in sec. \ref{section:scatter_data}.}
    \label{figure:oc_28}
\end{figure*}

We searched for the shrinking of the orbital period To constrain the orbital decay effects. Similar to \cite{patra}, \cite{yee} and \cite{turner} we implement two models to fit the timing data. The first one is the linear ephemeris model, which assumes that the orbital period is constant:
\begin{equation}
    T_\mathrm{lin} = T_{0} + P \times E
\end{equation}
where $E$, $T_{0}$ and $P$ are respectively the transit epoch, the reference mid-transit time (corresponding to $E=0$) and the orbital period. The second model is a quadratic one, which assumes that the orbital period is varying uniformly over time: 
\begin{equation}
    T_{quad}= T_{0} + P \times E + \frac{1}{2} \frac{\mathrm{d}P}{\mathrm{d}E} \times E^{2}
    \label{quadephe}
\end{equation}
where d$P$/d$E$ is the change in the orbital period between consecutive transits, which is assumed constant. The best fitting model parameters were found by performing a DE-MCMC analysis (100 000 steps with 1000 burn-in steps) of all the timings in Table~\ref{table:T0}. We adopted as the new zeroth epoch the transit time closest to the average of the available timings, to minimize the correlation between the ephemeris parameters, as done for instance by \citet{dragomir} and \citet{mallon}. 
The results of the quadratic model are:
\begin{equation}\label{eq:asiagofit}
\begin{split}
    &T_{0} = 2457280.093510 \pm  0.000067 & \textrm{(BJD-TDB),} \\
    &P = 1.0914192185 \pm 0.000000041 & \textrm{days,} \\
    &\frac{\mathrm{d}P}{\mathrm{\mathrm{d}}E} = (−1.063 \pm 0.093 ) \times 10^{-9} & \textrm{days/orbit}, \\
    &\frac{\mathrm{d}P}{\mathrm{d}t} = -30.72 \pm 2.67   & \textrm{ms/yr.}\\
\end{split}
\end{equation}
The corresponding $O-C $ (``Observed minus Calculated'') diagram is shown in Fig.~\ref{figure:oc_28} as a function of the epoch $E$. The derivative of the orbital period d$P$/d$E$ is derived from the quadratic coefficient of the best fitting parabola in Eq.~\ref{quadephe}, and then translated into d$P$/d$t$. Its negative sign indicates a decrease in the orbital period with a rate of 0.03071 seconds per year. The corresponding orbital decay timescale would be $\tau = -(dP/dt)/P$ = 3.07 $\pm$ 0.26 Myr, a value consistent with the results of \cite{turner} and \cite{wong} (See Table \ref{tab:confrontodP}).

\begin{figure*}[tb]
\sidecaption
    \centering
    \includegraphics[width=1.4\columnwidth]{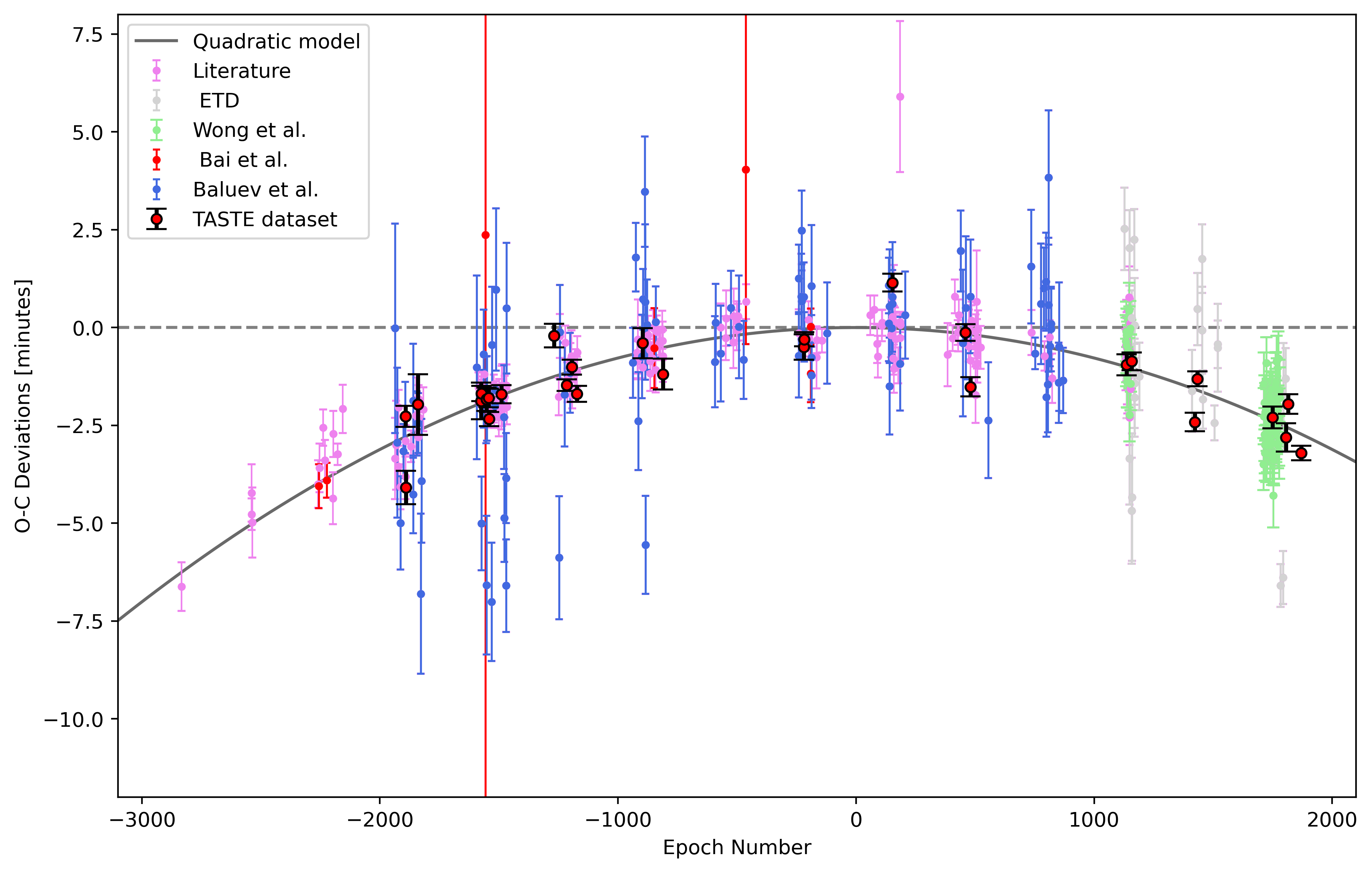}
    \caption{Observed minus calculated ($O-C$) diagram of WASP-12b for all the available transit timing data.  The reference ephemeris here is the linear part of Eq.~\ref{eq:litfit}, while the additional quadratic term (based on d$P$/d$E$) is plotted as a black solid line. 
    The Asiago data are plotted as black and red points, while the remaining points come from the literature:  \cite{bai} (red points, including TESS), \cite{wong} (light green), \cite{baluev} (light blue), other literature works (available at CDS, pink). Amateur timings from ETD are plotted in grey.}
    \label{figure:oc_total}
\end{figure*}

To further test the orbital decay hypothesis, we combine our homogeneous ground-based data set with all the available mid-transit times (356) compiled by \citealt{bai}. The data encompass timings from ground-based facilities as well as TESS results and data by amateur astronomers collected by ETD (\citealt{poddan}). All the literature timings and the filters used are listed in the additonal tables at CDS. The resulting parameters of the quadratic fit, performed as in the former case, are:
\begin{equation}\label{eq:litfit}
\begin{split}
    &T_{0} = 2457607.519305 \pm  0.000032 & \textrm{(BJD-TDB),} \\
    &P = 1.091418901 \pm 0.000000018 & \textrm{days,} \\
    &\frac{\mathrm{d}P}{\mathrm{d}E} = (-1.082 \pm 0.042 ) \times 10^{-9} & \textrm{days/orbit}, \\
    &\frac{\mathrm{d}P}{\mathrm{d}t} = -31.27 \pm 1.23  & \textrm{ms/yr.}\\
\end{split}
\end{equation}
We used these parameters to compute the $O-C $ diagram shown in Fig.~\ref{figure:oc_total}. We point out that the planetary parameters reported in Table~\ref{table:orbital_parameters_wasp12b} were derived based on Asiago photometry only, and are therefore independent from any literature work. 

As already shown by several previous studies \citep{patra2017,yee,turner} the quadratic model provides a much better fit with respect to the linear ephemeris.
This is also confirmed in our goodness-of-fit statistic, by the quadratic model (25 degrees of freedom, d.o.f) chi-square ($\mathrm{\chi^{2}}$) 770.87 against the linear model (26 d.o.f) value of 2463.66 and by the Bayesian Information Criterion (BIC; \citealt{gideon}). The orbital decay model is favored with a value of $\mathrm{\Delta}$(BIC) = 1693 against the linear model.

\section{Discussion}
\label{section:discussion}

\subsection{Tidal Orbital Decay}
The measured fast rate of orbital decay for WASP-12b has been attributed to tidal dissipation inside its host star. A short-period planetary system experiences a tidal orbital decay if the host star rotates slower than the orbital period of the planet. In this scenario, the orbital angular momentum of the planet is transferred to the spin of the star because the tidal bulge lags behind the position of the planet in its orbit. The orbital energy of the planet, deposited in the tidal bulge within the host star, is dissipated by tides, causing the star to spin up (\citealt{zahn}; \citealt{hut}; \citealt{eggleton}). The transfer of angular momentum from the planetary orbit to the stellar spin causes the orbital separation between the planet and the host star to shrink --along with orbital circularization-- leading ultimately to the engulfment of the planet after the stellar Roche radius limit is reached (\citealt{levrard}).
Orbital decay in a close-in, short-period exoplanet can be detected via long-monitoring transit timing variations observational campaigns (\citealt{miralda}; \citealt{agol1}; \citealt{holman}).

\subsection{Stellar tidal quality factor}
\label{section:quality_factor}
The tidal quality factor $Q$ is defined as the ratio between the total energy of the tide and the total energy dissipated in one tidal period \citep[see, e.g. Eq. 2.19 of][]{zahn08}.
When dealing with tidal dissipation, we pair $Q$ with the Love number $k_{2}$, which takes into account the internal density stratification of the object.
Directly measuring the decrease of the orbital period can be used to estimate such a so-called modified stellar tidal quality factor $Q^{\prime}_{\ast} \equiv 3Q_{\ast}/2k_{2}$ (\citealt{Matsumura_2010}; \citealt{Hoyer_2016}; \citealt{penev}; \citealt{turner22}). This dimensionless parameter measures the efficiency of the dissipation of the kinetic energy of the tides inside the star. We can deduce its value using the constant-phase-lag (CPL) model of \cite{GOLDREICH}:
\begin{equation}
    Q^{\prime}_{\ast} = - \frac{27}{2} \pi \left (\frac{M_{p}}{M_{\star}} \right ) \left (\frac{a}{R_{\star}} \right )^{-5} \left (\frac{\mathrm{d}P}{\mathrm{d}t} \right )^{-1}
\end{equation}
where $M_{p}$/$M_{\star}$ is the planet-to-star mass ratio, $R_{\star}$ is the stellar radius, $a$ is the semi- major axis and d$P$/d$t$ is the period derivative.
By taking our derived values of $M_{\star}$, $dP/dt$, $R_{\star}$, $a$, see Sect. \ref{section: stellaparameters} and Sect. \ref{timing_analysis}, and the planetary mass of \cite{Chakrabarty}, we obtained a modified tidal quality factor of
\begin{equation}
    Q^{\prime}_{\ast}= (2.13 \pm 0.18) \times 10^{5}.
\end{equation}
The planet mass is assumed to be constant, which we know from \cite{lai} not to be the case. This value is slightly higher, but still consistent, compared to what is derived by \cite{turner} and \cite{wong}.

In the  literature, there is a substantial disagreement between the value of $Q^{\prime}_{\ast}$ determined from WASP-12's studies and the observed or theoretically predicted ranges of values.
Our derived value is at the lower limit of the observed ranges as derived from population studies of hot Jupiters ($10^{5.5}$-$10^{6.5}$; \citealt{jackson}; \citealt{husnoo}; \citealt{bonomo}; \citealt{barker}) and binary star systems ($10^{5}$-$10^{7}$;  \citealt{meibom}; \citealt{olgivelin}; \citealt{lanza11}; \citealt{meibom15}). \cite{collier18} used a form of Bayesian hierarchical inference to numerically estimate the modified stellar quality factor for hot Jupiters. Their analysis found mean values of $Q^{\prime}_{\ast}$ $\sim$ $10^{7.3}$ and $10^{8.3}$, for dynamical and equilibrium tide regimes, respectively. The authors expected a departure from a linear ephemeris for WASP-12b of less than 1 s, on a 20-yr baseline. We can see from Fig.\ref{figure:oc_28} and Fig.\ref{figure:oc_total} that our observations greatly exceed that prediction  showing a variation of minutes on a 12-yr baseline. \cite{hamer&schlaufam} and \cite{barker} argued that the findings of population-wide studies of hot Jupiters
should be taken as  approximate estimates considering that  the value of $Q^{\prime}_{\ast}$ depends on the physical and mechanical properties (orbital period, stellar interior structure and rotation period, planet and stellar mass) of the individual star-planet systems. In this regard, \cite{rosario} studied WASP-18b, a massive hot Jupiter ($\sim$11~$M_{j}$) on a 0.94 days period orbiting a star with a spectral type and mass similar to those of WAPS-12, finding however a lower limit for the tidal quality factor of two orders of magnitude higher than that of WASP-12b.

Furthermore, based on the current understanding of tidal dissipation processes, the physical mechanism driving this efficient dissipation inside WASP-12 is yet to be understood. The tide in the star is frequently divided into two components: an equilibrium tide and a dynamical tide (\citealt{zahn}).
The damping of the equilibrium tide in the convective layers is too weak to explain the observed strong dissipation. Linear and non-linear wave-breaking of the dynamical tide (\textit{g}-modes) near the radiative core of the star would lead to a sufficiently  enhanced dissipation ($Q^{\prime}_{\ast}$ $\sim$ $10^{5}$) only if the star has entered its subgiant stage of evolution \citep{weinberg}.

The presence and size of the radiative core, which can be derived from the internal structure of the star, is fundamental to understand the tidal decay mechanism. If WASP-12 is still on the main sequence, it would possess a convective envelope and a small convective core. However if it has evolved to become a subgiant it would hold a radiative core, allowing gravity waves to deposit their angular momentum at the radiative-convective boundaries by breaking, resulting in a fast tidal decay (\citealt{weinberg}; \citealt{weinberg2023}).
Our precise observational constraints on the stellar parameters, such as mass, effective temperature, age  and metallicity, are however consistent with the main-sequence structure if compared to the theoretical models of \cite{weinberg} and \cite{bailey}.
In order to further investigate the evolutionary phase of WASP-12 we used the MESA Isochrone \& Stellar Tracks (MIST) stellar evolution models, to extract the theoretical isochrone equivalent to our derived value of [Fe/H] and stellar age (see Table. \ref{table:derived_stellar_parameters}).
We placed the star in the $\log(T_\mathrm{eff})$-$\log({L/L_{\odot}})$ diagram (H-R diagram), plotted in Fig \ref{figure:H-R}, where the isochrone is displayed as a solid black line.
We can notice that the star is located in an ambiguous position, between the main sequence and the turn-off, but far away from the subgiant branch.
This could tell us, as suggested by \cite{makarov}, that the star has entered the post-turn-off stage where hydrogen fusion in the core has ceased,ultimately leading to the core contraction and the burning of hydrogen in a shell around the core.

Recent studies proposed alternative hypotheses in which the observed tidal orbital decay can be explained by obliquity tides \citep{millholland}, planetary eccentricity tides \citep{makarov}, or a spin-orbit coupling mediated by a non-axisymmetric gravitational quadrupole moment inside the planet \citep{lanza20}.
Although significant improvements have been made in our observations of the system, there is still a need for more precise modeling to address the numerous unresolved questions and test both stellar evolution and tidal theories. A promising approach to determine whether WASP-12 is a main-sequence or a sub-giant star by measuring the age of the star, the mixing length parameter $\alpha$, the size of the convective core or the presence of a core rotation, is asteroseismology, as suggested by previous studies (\citealt{deheuvels}; \citealt{weinberg}; \citealt{bailey}; \citealt{fellay}).

\subsection{The search for orbital decay in other systems}
Among the population of hot Jupiters, WASP-12b is the first case where orbital decay has been detected. Several studies over the years investigated the decaying orbit of hot Jupiters, which possess orbital features similar to the target of our study (\citealt{patra}; \citealt{ivshina}; \citealt{hagey}).

\cite{harre_2023} used CHEOPS and TESS high-precision photometric data to investigate the orbital decay of WASP-4 b, KELT-9 b and KELT-16 b.
They found, in agreement with \cite{turner22} amd \cite{bouma}, a changing orbit only for WASP-4b, but even if the orbital decay model is preferred, apsidal precession cannot be ruled out with the available data.
Additional transit and occultations are needed to clearly identify the cause of the period variation.

\cite{Vissapragada_2022} presented strong evidence for the tidal decay of Kepler-1658 b, a close-in giant planet orbiting a sub-giant star, using a thirteen-year observational baseline. Using RVs, the authors ruled out apsidal precession on theoretical grounds and line-of-sight acceleration. The computed value of the modified tidal quality factors ($Q^{\prime}_{\ast} = 2.50^{+0.85}_{-0.62} \times 10^{4}$) fully agrees with what should result from the dissipation of inertial waves in the convective zone.
The recently discovered hot-Jupiter TOI-2109b \citep{Wong_2021} presents orbital and stellar characteristics that could lead to a stronger tidal dissipation than WASP-12b. Still new observations with a long observational baseline are needed to detect a potential orbital decay.

According to expected tidal decay rates and host star properties, many other planets should exhibit significant rates of orbital decay, e.g., WASP-18b \citep{collier18}, WASP-19b \citep{rosario}, WASP-43b \citep{patra}, WASP-103b  \citep{barros}, WASP-161b \citep{yang}, KELT-16b \citep{harre_2023}, HATS-18b \citep{southworth2022}. So far, not enough evidence has been found from the available data to confirm a decaying orbit for any of these systems.
Several effects can produce an $O-C$ variation over a long baseline, thus to have a successful validation of a planet tidal orbital inspiral and discard any other different possible cause it is necessary a decade-long observational baseline. 
Many of these targets will be re-observed by TESS, CHEOPS, PLATO \citep{nascimbeni_2022} and Ariel \citep{borsato_2022} in the near future and could be used in a future collective study. However, as advocated by \citet{hamer&schlaufam}, we additionally suggest that the focus of orbital decay studies should be on planets orbiting host stars at the end of their main-sequence lifetime.

\subsection{Excess scatter in the timing data}
\label{section:scatter_data}

An evident feature of our data is the presence of an excess scatter in the $O-C$ plot (Fig.~\ref{figure:oc_28}): the reduced $\chi^2$ of the fit on the Asiago data set is $\chi_\mathrm{r}^2=5.52$ ($\chi^2 \sim 138$ for 25 d.o.f), implying that the error bars are about $\sim 2.3$ times too small with respect to a well-behaved Gaussian distribution having the same variance. This cannot be attributed to issues in the absolute calibration of our time stamps, or to systematic errors left uncorrected by our photometric pipeline, as other TASTE studies carried out with the same setup and data analysis technique demonstrated a timing accuracy at the 1~s level or better \citep{BrownSevilla2021}. Such an excess is not unique to our unpublished data set: the best fit to the overall $O-C$ diagram (Fig.~\ref{figure:oc_total}) yields $\chi^2_\mathrm{r}\sim 2$, and other independent data sets reach $\chi^2_\mathrm{r}$ significantly larger than one (e.~g., $\chi^2_\mathrm{r}\sim 2.1$ for \citealt{baluev} and $\chi^2_\mathrm{r}\sim 8$ for ETD). 

We investigated the impact of correlated noise on our results by analyzing the residuals of our light curve, i.e. after the best-fit transit model has been subtracted. This was done by calculating the $\beta$ factor after binning the time series at different time scales (\citealp{gillon_2006}; \citealt{winn_2007a}). 
When correlated noise is present, the binned rms $\sigma$ is larger by a factor of $\beta$ with respect to the ideal white-noise case, where the rms is supposed to decrease by the square root of the number of points per bin ($\sqrt{N}$). At the time scale corresponding to the ingress/egress duration ($\sim$23 min for WASP-12b), i.e., the most relevant time scale for the transit time estimation, $\beta$ usually lies in the 1.5-2 range for well-behaved ground-based photometry \citep{winn_2007b}. Among our 28 light curves, only four of them (\texttt{\#2}, \texttt{7}, \texttt{20}, \texttt{28}) show an unusually large amount of red noise ($\beta\gg2$), but \texttt{\#20} is the only one also being a significant outlier in the $O-C$ diagram. 

As a further check, we binned each complete light curve at the time scale of the ingress duration (0.0159 days), then re-did the global fitting process on complete transits in the exact same way as explained in Section~\ref{section_fitting}. The resulting $\chi^2_\mathrm{r}$ is 3.2 ($\chi^2 =64.8$ with 20~d.~o.~f.), that is, still considerably larger than one. Even though the error bars increased, the sign of the most significant outliers is unchanged. We also emphasize that our best-fit $\mathrm{d}P/\mathrm{d}t$ in the binned case is $-29\pm4$~msec/year, fully in agreement with the previous value of $-30.7\pm2.7$~msec/year obtained without binning the light curves (Eq.~\ref{eq:asiagofit}).

These exercises demonstrate that, while a few light curves show a higher-than-usual amount of correlated noise, there is no obvious link with the excess scatter we found in the measured transit times. We also emphasize that the presence of partial transits in our data set cannot explain said excess: the reduced $\chi ^2$ for just our 22 complete transits is $\chi^2_r=5.78$ ($\chi^2=110$ for 19 d.~o.~f.), almost identical to the $\chi^2_r=5.52$ value for the full distribution ($\chi^2=137$ for 25 d.~o.~f.).

It is well known in the literature that comparing timing data extracted from ground-based transit light curves can be prone to systematic errors due to a combination of telluric/instrumental red noise and different reduction techniques \citep{baluev}. On the other hand, excess scatter is sometimes noticed also when dealing with space-based data (as in \citealt{turner22,harre_2023} in the case of WASP-4), and it is usually attributed to the presence of stellar activity. Little is known about the stellar activity properties of WASP-12b, since its anomalously low $R_{HK}$ value could be due to the presence of extra absorption from escaped circumstellar gas \citep{fossati} and no coherent rotational modulation can be detected from its light curve. Yet, significant bumps are visible in the residuals of the TESS data \citep{wong}, and at least one of our light curves (\texttt{\#23}) shows an in-transit anomaly which is difficult to explain with systematic errors (Section~\ref{section:data_analysis}), suggesting that in some cases inhomogeneities may arise on the stellar photosphere of WASP-12 but do not result in a detectable rotational modulation because of the high inclination of the stellar axis with respect to the observer, also suggest by the unusually low $v\sin i$. If the pole-on hypothesis is true, the timing scatter we see could be the only observable hint so far that WASP-12 is actually a significantly active star.

Intriguingly, even though running a GLS periodogram on our $O-C$ residuals does not result in any significant frequency, the excess scatter appears to be a time-dependent phenomenon. Transits gathered during some observing seasons (such as 2011-2012 and 2019-2020) line up within seconds from the average quadratic ephemeris and return $\chi_\mathrm{r}^2\sim 1$ even on the Asiago subset, while others (such as 2020-2021) appears much more scattered on different and independent data sets.

Forthcoming  additional photometry from TESS (scheduled on two sectors --71 and 72-- in Fall 2023) will possibly shed some more light on this behaviour. Meanwhile, we emphasize that, despite the scatter excess, our homogeneous data set, published here for the first time, represents a fully independent confirmation of the tidal decay process in WASP-12b, at an estimated rate perfectly consistent with those derived by previous studies.
\section{Summary and Conclusions}
\label{section:summary}
In this paper we inspected the tidal orbital decay of the Ultra-hot Jupiter WASP-12b, using 28 unpublished, high-precision, ground-based light curves collected over twelve years, from 2010 to 2022, as part of the TASTE program. 
We applied a DE-MCMC approach, for a linear (constant period) and quadratic model (orbital decay), to the timing of our homogeneously collected observations confirming the decaying orbit of WASP-12b.

Our new planetary and transit parameters are in good agreement with previous studies. However, combining all the timing data from literature we found some discrepancies in the $O-C$ plot.
The timing data shows an excess of scatter that cannot be attributed to errors in the photometric pipeline or the calibration of the time stamps.
This raises several questions which will require further study to resolve.

We used HARPS-N observations to investigate the evolutionary stage of the host star, which is fundamental to explain the high rate of tidal dissipation.
The applied spectroscopic observations allowed us to derive precise stellar parameters of the host star, however, we did not arrive at a firm conclusion regarding the evolutionary stage. The stellar parameters enable us to place the star in the main sequence close to the turn-off point and away from the subgiant branch. However, the fast tidal dissipation requires a subgiant scenario.

\begin{table*}[]
    \centering\small\centering\renewcommand{\arraystretch}{1.4}
        \caption{Comparison of the values of period change rate and orbital decay timescale of WASP-12b as estimated by different literature works.}
    \begin{tabular}{l c c}
    \hline\hline
         Reference & Period change rate d$P$/d$t$ & Orbital decay timescale $P$ / (d$P$/d$t$) \\
         & [ms/year] & [Myr] \\
         \hline
         This work & $-$30.71 $\pm$ 2.67 & 3.07 $\pm$ 0.26 \\
         \citealt{patra2017} & $-$29 $\pm$ 3 & 3.2 $\pm$ 0.38$^\dagger$ \\ 
         \citealt{yee} &  $-$29 $\pm$ 2 & 3.25 $\pm$ 0.24\\
         \citealt{turner} & $-$32.53 $\pm$ 1.6 & 2.90 $\pm$ 0.14 \\
         \citealt{ivshina} & $-$32.53 $\pm$ 1.62 & 3.11 $\pm$ 0.11 $^\dagger$ \\
         \citealt{wong} & $-$29.81 $\pm$ 0.94 & 3.16 $\pm$ 0.10  \\
         \citealt{bai} & $-$37.14 $\pm$ 1.31  & 2.54 $\pm$ 0.09 \\
         \hline
    \end{tabular}
\tablefoot{$\dagger$ value calculated by us, missing in the original paper.}
    \label{tab:confrontodP}
\end{table*}

\begin{acknowledgements}
This publication was produced while attending the PhD program in Space Science and Technology at the University of Trento, Cycle XXXVIII, with the support of a scholarship co-financed by the Ministerial Decree no. 351 of 9th April 2022, based on the NRRP - funded by the European Union - NextGenerationEU - Mission 4 "Education and Research", Component 2 "From Research to Business", Investment 3.3.
Based on observations collected at Copernico 1.82 m telescope and at the Schmidt 67/92 telescope.
We thank the GAPS collaboration for kindly providing the spectra of WASP-12. This research has made use of the SIMBAD database (operated at CDS, Strasbourg, France;\citealt{Wenger2000}), the VARTOOLS Light Curve Analysis Program (version 1.39 released October 30, 2020, \citealt{Hartman_and_Bakos_2016}), TOPCAT and STILTS \citep{Taylor2005,Taylor2006}, NASAs Astrophysics Data System (ADS) bibliographic services.
LBo, VNa, and GPi acknowledge support
from CHEOPS ASI-INAF agreement n. 2019-29-HH.0.
V.N., G.P., L.M. acknowledge financial support from the Bando Ricerca Fondamentale INAF 2023, Data Analysis Grant: "Characterization of transiting exoplanets by exploiting the unique synergy between TASTE and TESS".
\end{acknowledgements}

\bibliographystyle{aa}
\bibliography{references}

\begin{thebibliography}{139}
\expandafter\ifx\csname natexlab\endcsname\relax\def\natexlab#1{#1}\fi

\bibitem[{{Agol} {et~al.}(2005){Agol}, {Steffen}, {Sari}, \&
  {Clarkson}}]{agol1}
{Agol}, E., {Steffen}, J., {Sari}, R., \& {Clarkson}, W. 2005, \mnras, 359, 567

\bibitem[{{Albrecht} {et~al.}(2012){Albrecht}, {Winn}, {Johnson}, {Howard},
  {Marcy}, {Butler}, {Arriagada}, {Crane}, {Shectman}, {Thompson}, {Hirano},
  {Bakos}, \& {Hartman}}]{albrecht}
{Albrecht}, S., {Winn}, J.~N., {Johnson}, J.~A., {et~al.} 2012, \apj, 757, 18

\bibitem[{{Antonetti} \& {Goodman}(2022)}]{Antonetti&Goodman_2022}
{Antonetti}, V. \& {Goodman}, J. 2022, \apj, 939, 91

\bibitem[{{Antoniciello} {et~al.}(2021){Antoniciello}, {Borsato}, {Lacedelli},
  {Nascimbeni}, {Barrag{\'a}n}, \& {Claudi}}]{antoniciello}
{Antoniciello}, G., {Borsato}, L., {Lacedelli}, G., {et~al.} 2021, \mnras, 505,
  1567

\bibitem[{Bai {et~al.}(2022)Bai, Gu, Wang, Sun, Kwok, \& Hui}]{bai}
Bai, L., Gu, S., Wang, X., {et~al.} 2022, Monthly Notices of the Royal
  Astronomical Society, 512, 3113

\bibitem[{{Bailey} \& {Goodman}(2019)}]{bailey}
{Bailey}, A. \& {Goodman}, J. 2019, \mnras, 482, 1872

\bibitem[{Baluev {et~al.}(2019)Baluev, Sokov, Jones, Shaidulin, Sokova,
  Nielsen, Benni, Schneiter, Villarreal D’Angelo, Fernández-Lajús,
  Di Sisto, Baştürk, Bretton, Wunsche, Hentunen, Shadick, Jongen, Kang, Kim,
  Pakštienė, Qvam, Knight, Guerra, Marchini, Salvaggio, Papini, Evans,
  Salisbury, Garcia, Molina, Garlitz, Esseiva, Ogmen, Karavaev, Rusov,
  Ibrahimov, \& Karimov}]{baluev}
Baluev, R.~V., Sokov, E.~N., Jones, H. R.~A., {et~al.} 2019, Monthly Notices of
  the Royal Astronomical Society, 490, 1294

\bibitem[{{Barker}(2020)}]{barker}
{Barker}, A.~J. 2020, \mnras, 498, 2270

\bibitem[{{Barros} {et~al.}(2022){Barros}, {Akinsanmi}, {Bou{\'e}}, {Smith},
  {Laskar}, {Ulmer-Moll}, {Lillo-Box}, {Queloz}, {Cameron}, {Sousa},
  {Ehrenreich}, {Hooton}, {Bruno}, {Demory}, {Correia}, {Demangeon}, {Wilson},
  {Bonfanti}, {Hoyer}, {Alibert}, {Alonso}, {Escud{\'e}}, {Barbato},
  {B{\'a}rczy}, {Barrado}, {Baumjohann}, {Beck}, {Beck}, {Benz}, {Bergomi},
  {Billot}, {Bonfils}, {Bouchy}, {Brandeker}, {Broeg}, {Cabrera}, {Cessa},
  {Charnoz}, {Damme}, {Davies}, {Deleuil}, {Deline}, {Delrez}, {Erikson},
  {Fortier}, {Fossati}, {Fridlund}, {Gandolfi}, {Mu{\~n}oz}, {Gillon},
  {G{\"u}del}, {Isaak}, {Heng}, {Kiss}, {des Etangs}, {Lendl}, {Lovis},
  {Magrin}, {Nascimbeni}, {Maxted}, {Olofsson}, {Ottensamer}, {Pagano},
  {Pall{\'e}}, {Parviainen}, {Peter}, {Piotto}, {Pollacco}, {Ragazzoni},
  {Rando}, {Rauer}, {Ribas}, {Santos}, {Scandariato}, {S{\'e}gransan}, {Simon},
  {Steller}, {Szab{\'o}}, {Thomas}, {Udry}, {Ulmer}, {Van Grootel}, \&
  {Walton}}]{barros}
{Barros}, S.~C.~C., {Akinsanmi}, B., {Bou{\'e}}, G., {et~al.} 2022, \aap, 657,
  A52

\bibitem[{Batalha {et~al.}(2013)Batalha, Rowe, Bryson, Barclay, Burke,
  Caldwell, Christiansen, Mullally, Thompson, Brown, Dupree, Fabrycky, Ford,
  Fortney, Gilliland, Isaacson, Latham, Marcy, Quinn, Ragozzine, Shporer,
  Borucki, Ciardi, Gautier, Haas, Jenkins, Koch, Lissauer, Rapin, Basri, Boss,
  Buchhave, Carter, Charbonneau, Christensen-Dalsgaard, Clarke, Cochran,
  Demory, Desert, Devore, Doyle, Esquerdo, Everett, Fressin, Geary, Girouard,
  Gould, Hall, Holman, Howard, Howell, Ibrahim, Kinemuchi, Kjeldsen, Klaus, Li,
  Lucas, Meibom, Morris, Pr{\v{s}}a, Quintana, Sanderfer, Sasselov, Seader,
  Smith, Steffen, Still, Stumpe, Tarter, Tenenbaum, Torres, Twicken, Uddin,
  Cleve, Walkowicz, \& Welsh}]{batalha}
Batalha, N.~M., Rowe, J.~F., Bryson, S.~T., {et~al.} 2013, 204, 24

\bibitem[{{Bell} {et~al.}(2019){Bell}, {Zhang}, {Cubillos}, {Dang}, {Fossati},
  {Todorov}, {Cowan}, {Deming}, {Zellem}, {Stevenson}, {Crossfield},
  {Dobbs-Dixon}, {Fortney}, {Knutson}, \& {Line}}]{bell2019}
{Bell}, T.~J., {Zhang}, M., {Cubillos}, P.~E., {et~al.} 2019, \mnras, 489, 1995

\bibitem[{{Biazzo} {et~al.}(2022){Biazzo}, {D'Orazi}, {Desidera}, {Turrini},
  {Benatti}, {Gratton}, {Magrini}, {Sozzetti}, {Baratella}, {Bonomo}, {Borsa},
  {Claudi}, {Covino}, {Damasso}, {Di Mauro}, {Lanza}, {Maggio}, {Malavolta},
  {Maldonado}, {Marzari}, {Micela}, {Poretti}, {Vitello}, {Affer}, {Bignamini},
  {Carleo}, {Cosentino}, {Fiorenzano}, {Giacobbe}, {Harutyunyan}, {Leto},
  {Mancini}, {Molinari}, {Molinaro}, {Nardiello}, {Nascimbeni}, {Pagano},
  {Pedani}, {Piotto}, {Rainer}, \& {Scandariato}}]{biazzo2022}
{Biazzo}, K., {D'Orazi}, V., {Desidera}, S., {et~al.} 2022, \aap, 664, A161

\bibitem[{{Biazzo} {et~al.}(2015){Biazzo}, {Gratton}, {Desidera}, {Lucatello},
  {Sozzetti}, {Bonomo}, {Damasso}, {Gandolfi}, {Affer}, {Boccato}, {Borsa},
  {Claudi}, {Cosentino}, {Covino}, {Knapic}, {Lanza}, {Maldonado}, {Marzari},
  {Micela}, {Molaro}, {Pagano}, {Pedani}, {Pillitteri}, {Piotto}, {Poretti},
  {Rainer}, {Santos}, {Scandariato}, \& {Zanmar Sanchez}}]{Biazzoetal2015}
{Biazzo}, K., {Gratton}, R., {Desidera}, S., {et~al.} 2015, \aap, 583, A135

\bibitem[{{Bonomo} {et~al.}(2017){Bonomo}, {Desidera}, {Benatti}, {Borsa},
  {Crespi}, {Damasso}, {Lanza}, {Sozzetti}, {Lodato}, {Marzari}, {Boccato},
  {Claudi}, {Cosentino}, {Covino}, {Gratton}, {Maggio}, {Micela}, {Molinari},
  {Pagano}, {Piotto}, {Poretti}, {Smareglia}, {Affer}, {Biazzo}, {Bignamini},
  {Esposito}, {Giacobbe}, {H{\'e}brard}, {Malavolta}, {Maldonado}, {Mancini},
  {Martinez Fiorenzano}, {Masiero}, {Nascimbeni}, {Pedani}, {Rainer}, \&
  {Scandariato}}]{bonomo}
{Bonomo}, A.~S., {Desidera}, S., {Benatti}, S., {et~al.} 2017, \aap, 602, A107

\bibitem[{{Borsato} {et~al.}(2022){Borsato}, {Nascimbeni}, {Piotto}, \&
  {Szab{\'o}}}]{borsato_2022}
{Borsato}, L., {Nascimbeni}, V., {Piotto}, G., \& {Szab{\'o}}, G. 2022,
  Experimental Astronomy, 53, 635

\bibitem[{{Bouma} {et~al.}(2019){Bouma}, {Winn}, {Baxter}, {Bhatti}, {Dai},
  {Daylan}, {D{\'e}sert}, {Hill}, {Kane}, {Stassun}, {Villasenor}, {Ricker},
  {Vanderspek}, {Latham}, {Seager}, {Jenkins}, {Berta-Thompson}, {Col{\'o}n},
  {Fausnaugh}, {Glidden}, {Guerrero}, {Rodriguez}, {Twicken}, \&
  {Wohler}}]{bouma}
{Bouma}, L.~G., {Winn}, J.~N., {Baxter}, C., {et~al.} 2019, \aj, 157, 217

\bibitem[{{Brown-Sevilla} {et~al.}(2021){Brown-Sevilla}, {Nascimbeni},
  {Borsato}, {Tartaglia}, {Nardiello}, {Granata}, {Libralato}, {Damasso},
  {Piotto}, {Pollacco}, {West}, {Colombo}, {Cunial}, {Piazza}, \&
  {Scaggiante}}]{BrownSevilla2021}
{Brown-Sevilla}, S.~B., {Nascimbeni}, V., {Borsato}, L., {et~al.} 2021, \mnras,
  506, 2122

\bibitem[{{Castelli} \& {Kurucz}(2003)}]{CastelliKurucz2003}
{Castelli}, F. \& {Kurucz}, R.~L. 2003, in Modelling of Stellar Atmospheres,
  ed. N.~{Piskunov}, W.~W. {Weiss}, \& D.~F. {Gray}, Vol. 210, A20

\bibitem[{{Chakrabarty} \& {Sengupta}(2019)}]{Chakrabarty}
{Chakrabarty}, A. \& {Sengupta}, S. 2019, \aj, 158, 39

\bibitem[{{Chan} {et~al.}(2011){Chan}, {Ingemyr}, {Winn}, {Holman},
  {Sanchis-Ojeda}, {Esquerdo}, \& {Everett}}]{chan}
{Chan}, T., {Ingemyr}, M., {Winn}, J.~N., {et~al.} 2011, \aj, 141, 179

\bibitem[{{Choi} {et~al.}(2016){Choi}, {Dotter}, {Conroy}, {Cantiello},
  {Paxton}, \& {Johnson}}]{choi}
{Choi}, J., {Dotter}, A., {Conroy}, C., {et~al.} 2016, \apj, 823, 102

\bibitem[{{Collier Cameron} \& {Jardine}(2018)}]{collier18}
{Collier Cameron}, A. \& {Jardine}, M. 2018, \mnras, 476, 2542

\bibitem[{Collins {et~al.}(2017)Collins, Kielkopf, \& Stassun}]{collins}
Collins, K.~A., Kielkopf, J.~F., \& Stassun, K.~G. 2017, 153, 78

\bibitem[{{Copperwheat} {et~al.}(2013){Copperwheat}, {Wheatley}, {Southworth},
  {Bento}, {Marsh}, {Dhillon}, {Fortney}, {Littlefair}, \&
  {Hickman}}]{copperweat}
{Copperwheat}, C.~M., {Wheatley}, P.~J., {Southworth}, J., {et~al.} 2013,
  \mnras, 434, 661

\bibitem[{{Cosentino} {et~al.}(2012){Cosentino}, {Lovis}, {Pepe}, {Collier
  Cameron}, {Latham}, {Molinari}, {Udry}, {Bezawada}, {Black}, {Born},
  {Buchschacher}, {Charbonneau}, {Figueira}, {Fleury}, {Galli}, {Gallie},
  {Gao}, {Ghedina}, {Gonzalez}, {Gonzalez}, {Guerra}, {Henry}, {Horne},
  {Hughes}, {Kelly}, {Lodi}, {Lunney}, {Maire}, {Mayor}, {Micela}, {Ordway},
  {Peacock}, {Phillips}, {Piotto}, {Pollacco}, {Queloz}, {Rice}, {Riverol},
  {Riverol}, {San Juan}, {Sasselov}, {Segransan}, {Sozzetti}, {Sosnowska},
  {Stobie}, {Szentgyorgyi}, {Vick}, \& {Weber}}]{cosentino2012}
{Cosentino}, R., {Lovis}, C., {Pepe}, F., {et~al.} 2012, in Society of
  Photo-Optical Instrumentation Engineers (SPIE) Conference Series, Vol. 8446,
  Ground-based and Airborne Instrumentation for Astronomy IV, ed. I.~S.
  {McLean}, S.~K. {Ramsay}, \& H.~{Takami}, 84461V

\bibitem[{{Cosentino} {et~al.}(2014){Cosentino}, {Lovis}, {Pepe}, {Collier
  Cameron}, {Latham}, {Molinari}, {Udry}, {Bezawada}, {Buchschacher},
  {Figueira}, {Fleury}, {Ghedina}, {Glenday}, {Gonzalez}, {Guerra}, {Henry},
  {Hughes}, {Maire}, {Motalebi}, \& {Phillips}}]{cosentino2014}
{Cosentino}, R., {Lovis}, C., {Pepe}, F., {et~al.} 2014, in Society of
  Photo-Optical Instrumentation Engineers (SPIE) Conference Series, Vol. 9147,
  Ground-based and Airborne Instrumentation for Astronomy V, ed. S.~K.
  {Ramsay}, I.~S. {McLean}, \& H.~{Takami}, 91478C

\bibitem[{{Covino} {et~al.}(2013){Covino}, {Esposito}, {Barbieri}, {Mancini},
  {Nascimbeni}, {Claudi}, {Desidera}, {Gratton}, {Lanza}, {Sozzetti}, {Biazzo},
  {Affer}, {Gandolfi}, {Munari}, {Pagano}, {Bonomo}, {Collier Cameron},
  {H{\'e}brard}, {Maggio}, {Messina}, {Micela}, {Molinari}, {Pepe}, {Piotto},
  {Ribas}, {Santos}, {Southworth}, {Shkolnik}, {Triaud}, {Bedin}, {Benatti},
  {Boccato}, {Bonavita}, {Borsa}, {Borsato}, {Brown}, {Carolo}, {Ciceri},
  {Cosentino}, {Damasso}, {Faedi}, {Mart{\'\i}nez Fiorenzano}, {Latham},
  {Lovis}, {Mordasini}, {Nikolov}, {Poretti}, {Rainer}, {Rebolo L{\'o}pez},
  {Scandariato}, {Silvotti}, {Smareglia}, {Alcal{\'a}}, {Cunial}, {Di
  Fabrizio}, {Di Mauro}, {Giacobbe}, {Granata}, {Harutyunyan}, {Knapic},
  {Lattanzi}, {Leto}, {Lodato}, {Malavolta}, {Marzari}, {Molinaro},
  {Nardiello}, {Pedani}, {Prisinzano}, \& {Turrini}}]{covino2013}
{Covino}, E., {Esposito}, M., {Barbieri}, M., {et~al.} 2013, \aap, 554, A28

\bibitem[{{Cowan} {et~al.}(2012){Cowan}, {Machalek}, {Croll}, {Shekhtman},
  {Burrows}, {Deming}, {Greene}, \& {Hora}}]{cowan}
{Cowan}, N.~B., {Machalek}, P., {Croll}, B., {et~al.} 2012, \apj, 747, 82

\bibitem[{{Cutri} {et~al.}(2003){Cutri}, {Skrutskie}, {van Dyk}, {Beichman},
  {Carpenter}, {Chester}, {Cambresy}, {Evans}, {Fowler}, {Gizis}, {Howard},
  {Huchra}, {Jarrett}, {Kopan}, {Kirkpatrick}, {Light}, {Marsh}, {McCallon},
  {Schneider}, {Stiening}, {Sykes}, {Weinberg}, {Wheaton}, {Wheelock}, \&
  {Zacarias}}]{cutri_2003}
{Cutri}, R.~M., {Skrutskie}, M.~F., {van Dyk}, S., {et~al.} 2003, VizieR Online
  Data Catalog, II/246

\bibitem[{{Dawson} \& {Johnson}(2018)}]{dawson}
{Dawson}, R.~I. \& {Johnson}, J.~A. 2018, \araa, 56, 175

\bibitem[{Debrecht {et~al.}(2018)Debrecht, Carroll-Nellenback, Frank, Fossati,
  Blackman, \& Dobbs-Dixon}]{debrecht}
Debrecht, A., Carroll-Nellenback, J., Frank, A., {et~al.} 2018, Monthly Notices
  of the Royal Astronomical Society, 478, 2592

\bibitem[{{Deheuvels} {et~al.}(2016){Deheuvels}, {Brand{\~a}o}, {Silva
  Aguirre}, {Ballot}, {Michel}, {Cunha}, {Lebreton}, \&
  {Appourchaux}}]{deheuvels}
{Deheuvels}, S., {Brand{\~a}o}, I., {Silva Aguirre}, V., {et~al.} 2016, \aap,
  589, A93

\bibitem[{{Dotter}(2016)}]{dotter16}
{Dotter}, A. 2016, \apjs, 222, 8

\bibitem[{{Dotter} {et~al.}(2008){Dotter}, {Chaboyer}, {Jevremovi{\'c}},
  {Kostov}, {Baron}, \& {Ferguson}}]{dotter08}
{Dotter}, A., {Chaboyer}, B., {Jevremovi{\'c}}, D., {et~al.} 2008, \apjs, 178,
  89

\bibitem[{{Doyle} {et~al.}(2014){Doyle}, {Davies}, {Smalley}, {Chaplin}, \&
  {Elsworth}}]{Doyleetal2014}
{Doyle}, A.~P., {Davies}, G.~R., {Smalley}, B., {Chaplin}, W.~J., \&
  {Elsworth}, Y. 2014, \mnras, 444, 3592

\bibitem[{{Dragomir} {et~al.}(2011){Dragomir}, {Kane}, {Pilyavsky},
  {Mahadevan}, {Ciardi}, {Gazak}, {Gelino}, {Payne}, {Rabus}, {Ramirez}, {von
  Braun}, {Wright}, \& {Wyatt}}]{dragomir}
{Dragomir}, D., {Kane}, S.~R., {Pilyavsky}, G., {et~al.} 2011, \aj, 142, 115

\bibitem[{Dressing \& Charbonneau(2013)}]{dressing}
Dressing, C.~D. \& Charbonneau, D. 2013, 767, 95

\bibitem[{{Eastman} {et~al.}(2010){Eastman}, {Siverd}, \& {Gaudi}}]{eastman}
{Eastman}, J., {Siverd}, R., \& {Gaudi}, B.~S. 2010, \pasp, 122, 935

\bibitem[{{Efroimsky} \& {Makarov}(2022)}]{makarov}
{Efroimsky}, M. \& {Makarov}, V.~V. 2022, Universe, 8, 211

\bibitem[{{Eggleton} {et~al.}(1998){Eggleton}, {Kiseleva}, \& {Hut}}]{eggleton}
{Eggleton}, P.~P., {Kiseleva}, L.~G., \& {Hut}, P. 1998, \apj, 499, 853

\bibitem[{{Fabrycky} \& {Tremaine}(2007)}]{fabrycky}
{Fabrycky}, D. \& {Tremaine}, S. 2007, \apj, 669, 1298

\bibitem[{{Fellay} {et~al.}(2023){Fellay}, {Pezzotti}, {Buldgen},
  {Eggenberger}, \& {Bolmont}}]{fellay}
{Fellay}, L., {Pezzotti}, C., {Buldgen}, G., {Eggenberger}, P., \& {Bolmont},
  E. 2023, \aap, 669, A2

\bibitem[{{Feroz} \& {Hobson}(2008)}]{feroz08}
{Feroz}, F. \& {Hobson}, M.~P. 2008, \mnras, 384, 449

\bibitem[{{Feroz} {et~al.}(2009){Feroz}, {Hobson}, \& {Bridges}}]{feroz09}
{Feroz}, F., {Hobson}, M.~P., \& {Bridges}, M. 2009, \mnras, 398, 1601

\bibitem[{{Feroz} {et~al.}(2019){Feroz}, {Hobson}, {Cameron}, \&
  {Pettitt}}]{feroz19}
{Feroz}, F., {Hobson}, M.~P., {Cameron}, E., \& {Pettitt}, A.~N. 2019, The Open
  Journal of Astrophysics, 2, 10

\bibitem[{{Foreman-Mackey} {et~al.}(2013){Foreman-Mackey}, {Hogg}, {Lang}, \&
  {Goodman}}]{foreman-mackey}
{Foreman-Mackey}, D., {Hogg}, D.~W., {Lang}, D., \& {Goodman}, J. 2013, \pasp,
  125, 306

\bibitem[{{Fortney} {et~al.}(2021){Fortney}, {Dawson}, \& {Komacek}}]{fortney}
{Fortney}, J.~J., {Dawson}, R.~I., \& {Komacek}, T.~D. 2021, Journal of
  Geophysical Research (Planets), 126, e06629

\bibitem[{Fossati {et~al.}(2010)Fossati, Haswell, Froning, Hebb, Holmes, Kolb,
  Helling, Carter, Wheatley, Cameron, Loeillet, Pollacco, Street, Stempels,
  Simpson, Udry, Joshi, West, Skillen, \& Wilson}]{fossati}
Fossati, L., Haswell, C.~A., Froning, C.~S., {et~al.} 2010, 714, L222

\bibitem[{{Gaia Collaboration}(2020)}]{gaia_eDR3_2020}
{Gaia Collaboration}. 2020, VizieR Online Data Catalog, I/350

\bibitem[{{Gillon} {et~al.}(2006){Gillon}, {Pont}, {Moutou}, {Bouchy},
  {Courbin}, {Sohy}, \& {Magain}}]{gillon_2006}
{Gillon}, M., {Pont}, F., {Moutou}, C., {et~al.} 2006, \aap, 459, 249

\bibitem[{Goldreich \& Soter(1966)}]{GOLDREICH}
Goldreich, P. \& Soter, S. 1966, Icarus, 5, 375

\bibitem[{{Granata} {et~al.}(2014){Granata}, {Nascimbeni}, {Piotto}, {Bedin},
  {Borsato}, {Cunial}, {Damasso}, \& {Malavolta }}]{granata_2014}
{Granata}, V., {Nascimbeni}, V., {Piotto}, G., {et~al.} 2014, Astronomische
  Nachrichten, 335, 797

\bibitem[{{Guilluy} {et~al.}(2022){Guilluy}, {Giacobbe}, {Carleo}, {Cubillos},
  {Sozzetti}, {Bonomo}, {Brogi}, {Gandhi}, {Fossati}, {Nascimbeni}, {Turrini},
  {Schisano}, {Borsa}, {Lanza}, {Mancini}, {Maggio}, {Malavolta}, {Micela},
  {Pino}, {Rainer}, {Bignamini}, {Claudi}, {Cosentino}, {Covino}, {Desidera},
  {Fiorenzano}, {Harutyunyan}, {Lorenzi}, {Knapic}, {Molinari}, {Pacetti},
  {Pagano}, {Pedani}, {Piotto}, \& {Poretti}}]{guilluy}
{Guilluy}, G., {Giacobbe}, P., {Carleo}, I., {et~al.} 2022, \aap, 665, A104

\bibitem[{{Hagey} {et~al.}(2022){Hagey}, {Edwards}, \& {Boley}}]{hagey}
{Hagey}, S.~R., {Edwards}, B., \& {Boley}, A.~C. 2022, \aj, 164, 220

\bibitem[{{Hamer} \& {Schlaufman}(2019)}]{hamer&schlaufam}
{Hamer}, J.~H. \& {Schlaufman}, K.~C. 2019, \aj, 158, 190

\bibitem[{{Harre} {et~al.}(2023){Harre}, {Smith}, {Barros}, {Bou{\'e}},
  {Csizmadia}, {Ehrenreich}, {Flor{\'e}n}, {Fortier}, {Maxted}, {Hooton},
  {Akinsanmi}, {Serrano}, {Ros{\'a}rio}, {Demory}, {Jones}, {Laskar},
  {Adibekyan}, {Alibert}, {Alonso}, {Anderson}, {Anglada}, {Asquier},
  {B{\'a}rczy}, {Barrado y Navascues}, {Baumjohann}, {Beck}, {Beck}, {Benz},
  {Billot}, {Biondi}, {Bonfanti}, {Bonfils}, {Brandeker}, {Broeg}, {Cabrera},
  {Cessa}, {Charnoz}, {Collier Cameron}, {Davies}, {Deleuil}, {Delrez},
  {Demangeon}, {Erikson}, {Fossati}, {Fridlund}, {Gandolfi}, {Gillon},
  {G{\"u}del}, {Hellier}, {Heng}, {Hoyer}, {Isaak}, {Kiss}, {Lecavelier des
  Etangs}, {Lendl}, {Lovis}, {Luntzer}, {Magrin}, {Nascimbeni}, {Olofsson},
  {Ottensamer}, {Pagano}, {Pall{\'e}}, {Persson}, {Peter}, {Piotto},
  {Pollacco}, {Queloz}, {Ragazzoni}, {Rando}, {Rauer}, {Ribas}, {Ricker},
  {Salmon}, {Santos}, {Scandariato}, {Seager}, {S{\'e}gransan}, {Simon},
  {Sousa}, {Steller}, {Szab{\'o}}, {Thomas}, {Udry}, {Ulmer}, {Van Grootel},
  {Walton}, {Wilson}, {Winn}, \& {Wohler}}]{harre_2023}
{Harre}, J.~V., {Smith}, A.~M.~S., {Barros}, S.~C.~C., {et~al.} 2023, \aap,
  669, A124

\bibitem[{{Hartman} \& {Bakos}(2016)}]{Hartman_and_Bakos_2016}
{Hartman}, J.~D. \& {Bakos}, G.~{\'A}. 2016, Astronomy and Computing, 17, 1

\bibitem[{Haswell {et~al.}(2012)Haswell, Fossati, Ayres, France, Froning,
  Holmes, Kolb, Busuttil, Street, Hebb, Cameron, Enoch, Burwitz, Rodriguez,
  West, Pollacco, Wheatley, \& Carter}]{haswell}
Haswell, C.~A., Fossati, L., Ayres, T., {et~al.} 2012, The Astrophysical
  Journal, 760, 79

\bibitem[{{Hebb} {et~al.}(2009){Hebb}, {Collier-Cameron}, {Loeillet},
  {Pollacco}, {H{\'e}brard}, {Street}, {Bouchy}, {Stempels}, {Moutou},
  {Simpson}, {Udry}, {Joshi}, {West}, {Skillen}, {Wilson}, {McDonald},
  {Gibson}, {Aigrain}, {Anderson}, {Benn}, {Christian}, {Enoch}, {Haswell},
  {Hellier}, {Horne}, {Irwin}, {Lister}, {Maxted}, {Mayor}, {Norton}, {Parley},
  {Pont}, {Queloz}, {Smalley}, \& {Wheatley}}]{hebb}
{Hebb}, L., {Collier-Cameron}, A., {Loeillet}, B., {et~al.} 2009, \apj, 693,
  1920

\bibitem[{{Himes} \& {Harrington}(2022)}]{Himes2022}
{Himes}, M.~D. \& {Harrington}, J. 2022, \apj, 931, 86

\bibitem[{{H{\o}g} {et~al.}(2000){H{\o}g}, {Fabricius}, {Makarov}, {Urban},
  {Corbin}, {Wycoff}, {Bastian}, {Schwekendiek}, \& {Wicenec}}]{tycho-2}
{H{\o}g}, E., {Fabricius}, C., {Makarov}, V.~V., {et~al.} 2000, \aap, 355, L27

\bibitem[{{Holman} \& {Murray}(2005)}]{holman}
{Holman}, M.~J. \& {Murray}, N.~W. 2005, Science, 307, 1288

\bibitem[{Howard {et~al.}(2012)Howard, Marcy, Bryson, Jenkins, Rowe, Batalha,
  Borucki, Koch, Dunham, Gautier, Cleve, Cochran, Latham, Lissauer, Torres,
  Brown, Gilliland, Buchhave, Caldwell, Christensen-Dalsgaard, Ciardi, Fressin,
  Haas, Howell, Kjeldsen, Seager, Rogers, Sasselov, Steffen, Basri,
  Charbonneau, Christiansen, Clarke, Dupree, Fabrycky, Fischer, Ford, Fortney,
  Tarter, Girouard, Holman, Johnson, Klaus, Machalek, Moorhead, Morehead,
  Ragozzine, Tenenbaum, Twicken, Quinn, Isaacson, Shporer, Lucas, Walkowicz,
  Welsh, Boss, Devore, Gould, Smith, Morris, Prsa, Morton, Still, Thompson,
  Mullally, Endl, \& MacQueen}]{howard}
Howard, A.~W., Marcy, G.~W., Bryson, S.~T., {et~al.} 2012, 201, 15

\bibitem[{Hoyer {et~al.}(2016)Hoyer, Pallé, Dragomir, \& Murgas}]{Hoyer_2016}
Hoyer, S., Pallé, E., Dragomir, D., \& Murgas, F. 2016, The Astronomical
  Journal, 151, 137

\bibitem[{{Husnoo} {et~al.}(2012){Husnoo}, {Pont}, {Mazeh}, {Fabrycky},
  {H{\'e}brard}, {Bouchy}, \& {Shporer}}]{husnoo}
{Husnoo}, N., {Pont}, F., {Mazeh}, T., {et~al.} 2012, \mnras, 422, 3151

\bibitem[{{Husser} {et~al.}(2013){Husser}, {Wende-von Berg}, {Dreizler},
  {Homeier}, {Reiners}, {Barman}, \& {Hauschildt}}]{husser}
{Husser}, T.~O., {Wende-von Berg}, S., {Dreizler}, S., {et~al.} 2013, \aap,
  553, A6

\bibitem[{{Hut}(1981)}]{hut}
{Hut}, P. 1981, \aap, 99, 126

\bibitem[{{Ivshina} \& {Winn}(2022)}]{ivshina}
{Ivshina}, E.~S. \& {Winn}, J.~N. 2022, \apjs, 259, 62

\bibitem[{{Jackson} {et~al.}(2008){Jackson}, {Greenberg}, \&
  {Barnes}}]{jackson}
{Jackson}, B., {Greenberg}, R., \& {Barnes}, R. 2008, \apj, 678, 1396

\bibitem[{{Jeffries} {et~al.}(2023){Jeffries}, {Jackson}, {Wright}, {Weaver},
  {Gilmore}, {Randich}, {Bragaglia}, {Korn}, {Smiljanic}, {Biazzo}, {Casey},
  {Frasca}, {Gonneau}, {Guiglion}, {Morbidelli}, {Prisinzano}, {Sacco},
  {Tautvai{\v{s}}ien{\.{e}}}, {Worley}, \& {Zaggia}}]{Jeffriesetal2023}
{Jeffries}, R.~D., {Jackson}, R.~J., {Wright}, N.~J., {et~al.} 2023, \mnras,
  523, 802

\bibitem[{Kipping(2013)}]{kipping2013}
Kipping, D.~M. 2013, Monthly Notices of the Royal Astronomical Society, 435,
  2152

\bibitem[{{Kreidberg}(2015)}]{kreidberg}
{Kreidberg}, L. 2015, \pasp, 127, 1161

\bibitem[{{Kreidberg} {et~al.}(2015){Kreidberg}, {Line}, {Bean}, {Stevenson},
  {D{\'e}sert}, {Madhusudhan}, {Fortney}, {Barstow}, {Henry}, {Williamson}, \&
  {Showman}}]{kreidberg15}
{Kreidberg}, L., {Line}, M.~R., {Bean}, J.~L., {et~al.} 2015, \apj, 814, 66

\bibitem[{{Lacedelli} {et~al.}(2021){Lacedelli}, {Malavolta}, {Borsato},
  {Piotto}, {Nardiello}, {Mortier}, {Stalport}, {Collier Cameron}, {Poretti},
  {Buchhave}, {L{\'o}pez-Morales}, {Nascimbeni}, {Wilson}, {Udry}, {Latham},
  {Bonomo}, {Damasso}, {Dumusque}, {Jenkins}, {Lovis}, {Rice}, {Sasselov},
  {Winn}, {Andreuzzi}, {Cosentino}, {Charbonneau}, {Di Fabrizio}, {Martnez
  Fiorenzano}, {Ghedina}, {Harutyunyan}, {Lienhard}, {Micela}, {Molinari},
  {Pagano}, {Pepe}, {Phillips}, {Pinamonti}, {Ricker}, {Scandariato},
  {Sozzetti}, \& {Watson}}]{lacedelli21}
{Lacedelli}, G., {Malavolta}, L., {Borsato}, L., {et~al.} 2021, \mnras, 501,
  4148

\bibitem[{{Lai} {et~al.}(2010){Lai}, {Helling}, \& {van den Heuvel}}]{lai}
{Lai}, D., {Helling}, C., \& {van den Heuvel}, E.~P.~J. 2010, \apj, 721, 923

\bibitem[{Lanza(2020)}]{lanza20}
Lanza, A.~F. 2020, Monthly Notices of the Royal Astronomical Society, 497, 3911

\bibitem[{{Lanza} {et~al.}(2011){Lanza}, {Damiani}, \& {Gandolfi}}]{lanza11}
{Lanza}, A.~F., {Damiani}, C., \& {Gandolfi}, D. 2011, \aap, 529, A50

\bibitem[{{Levrard} {et~al.}(2009){Levrard}, {Winisdoerffer}, \&
  {Chabrier}}]{levrard}
{Levrard}, B., {Winisdoerffer}, C., \& {Chabrier}, G. 2009, \apjl, 692, L9

\bibitem[{{Li} {et~al.}(2010){Li}, {Miller}, {Lin}, \& {Fortney}}]{li}
{Li}, S.-L., {Miller}, N., {Lin}, D. N.~C., \& {Fortney}, J.~J. 2010, \nat,
  463, 1054

\bibitem[{{Lin} {et~al.}(1996){Lin}, {Bodenheimer}, \& {Richardson}}]{lin}
{Lin}, D.~N.~C., {Bodenheimer}, P., \& {Richardson}, D.~C. 1996, \nat, 380, 606

\bibitem[{{Lind} {et~al.}(2009){Lind}, {Asplund}, \& {Barklem}}]{Lindetal2009}
{Lind}, K., {Asplund}, M., \& {Barklem}, P.~S. 2009, \aap, 503, 541

\bibitem[{{Maciejewski} {et~al.}(2016){Maciejewski}, {Dimitrov},
  {Fern{\'a}ndez}, {Sota}, {Nowak}, {Ohlert}, {Nikolov}, {Bukowiecki}, {Hinse},
  {Pall{\'e}}, {Tingley}, {Kjurkchieva}, {Lee}, \& {Lee}}]{mac2016}
{Maciejewski}, G., {Dimitrov}, D., {Fern{\'a}ndez}, M., {et~al.} 2016, \aap,
  588, L6

\bibitem[{{Maciejewski} {et~al.}(2013){Maciejewski}, {Dimitrov}, {Seeliger},
  {Raetz}, {Bukowiecki}, {Kitze}, {Errmann}, {Nowak}, {Niedzielski}, {Popov},
  {Marka}, {Go{\'z}dziewski}, {Neuh{\"a}user}, {Ohlert}, {Hinse}, {Lee}, {Lee},
  {Yoon}, {Berndt}, {Gilbert}, {Ginski}, {Hohle}, {Mugrauer}, {R{\"o}ll},
  {Schmidt}, {Tetzlaff}, {Mancini}, {Southworth}, {Dall'Ora}, {Ciceri},
  {Zambelli}, {Corfini}, {Takahashi}, {Tachihara}, {Benk{\H{o}}},
  {S{\'a}rneczky}, {Szabo}, {Varga}, {Va{\v{n}}ko}, {Joshi}, \&
  {Chen}}]{mac2013}
{Maciejewski}, G., {Dimitrov}, D., {Seeliger}, M., {et~al.} 2013, \aap, 551,
  A108

\bibitem[{{Maciejewski} {et~al.}(2011){Maciejewski}, {Errmann}, {Raetz},
  {Seeliger}, {Spaleniak}, \& {Neuh{\"a}user}}]{macie2011}
{Maciejewski}, G., {Errmann}, R., {Raetz}, S., {et~al.} 2011, \aap, 528, A65

\bibitem[{{Maciejewski} {et~al.}(2018){Maciejewski}, {Fern{\'a}ndez},
  {Aceituno}, {Mart{\'\i}n-Ruiz}, {Ohlert}, {Dimitrov}, {Szyszka}, {von Essen},
  {Mugrauer}, {Bischoff}, {Michel}, {Mallonn}, {Stangret}, \&
  {Mo{\'z}dzierski}}]{mac2018}
{Maciejewski}, G., {Fern{\'a}ndez}, M., {Aceituno}, F., {et~al.} 2018, \actaa,
  68, 371

\bibitem[{{Malavolta} {et~al.}(2018){Malavolta}, {Mayo}, {Louden}, {Rajpaul},
  {Bonomo}, {Buchhave}, {Kreidberg}, {Kristiansen}, {Lopez-Morales}, {Mortier},
  {Vanderburg}, {Coffinet}, {Ehrenreich}, {Lovis}, {Bouchy}, {Charbonneau},
  {Ciardi}, {Collier Cameron}, {Cosentino}, {Crossfield}, {Damasso},
  {Dressing}, {Dumusque}, {Everett}, {Figueira}, {Fiorenzano}, {Gonzales},
  {Haywood}, {Harutyunyan}, {Hirsch}, {Howell}, {Johnson}, {Latham}, {Lopez},
  {Mayor}, {Micela}, {Molinari}, {Nascimbeni}, {Pepe}, {Phillips}, {Piotto},
  {Rice}, {Sasselov}, {S{\'e}gransan}, {Sozzetti}, {Udry}, \&
  {Watson}}]{malavolta2018}
{Malavolta}, L., {Mayo}, A.~W., {Louden}, T., {et~al.} 2018, \aj, 155, 107

\bibitem[{{Malavolta} {et~al.}(2016){Malavolta}, {Nascimbeni}, {Piotto},
  {Quinn}, {Borsato}, {Granata}, {Bonomo}, {Marzari}, {Bedin}, {Rainer},
  {Desidera}, {Lanza}, {Poretti}, {Sozzetti}, {White}, {Latham}, {Cunial},
  {Libralato}, {Nardiello}, {Boccato}, {Claudi}, {Cosentino}, {Covino},
  {Gratton}, {Maggio}, {Micela}, {Molinari}, {Pagano}, {Smareglia}, {Affer},
  {Andreuzzi}, {Aparicio}, {Benatti}, {Bignamini}, {Borsa}, {Damasso}, {Di
  Fabrizio}, {Harutyunyan}, {Esposito}, {Fiorenzano}, {Gandolfi}, {Giacobbe},
  {Gonz{\'a}lez Hern{\'a}ndez}, {Maldonado}, {Masiero}, {Molinaro}, {Pedani},
  \& {Scandariato}}]{malavolta2016}
{Malavolta}, L., {Nascimbeni}, V., {Piotto}, G., {et~al.} 2016, \aap, 588, A118

\bibitem[{{Mallonn} {et~al.}(2019){Mallonn}, {von Essen}, {Herrero},
  {Alexoudi}, {Granzer}, {Sosa}, {Strassmeier}, {Bakos}, {Bayliss}, {Brahm},
  {Bretton}, {Campos}, {Carone}, {Col{\'o}n}, {Dale}, {Dragomir}, {Espinoza},
  {Evans}, {Garcia}, {Gu}, {Guerra}, {Jongen}, {Jord{\'a}n}, {Kang}, {Keles},
  {Kim}, {Lendl}, {Molina}, {Salisbury}, {Scaggiante}, {Shporer}, {Siverd},
  {Sokov}, {Sokova}, \& {W{\"u}nsche}}]{mallon}
{Mallonn}, M., {von Essen}, C., {Herrero}, E., {et~al.} 2019, \aap, 622, A81

\bibitem[{Matsumura {et~al.}(2010)Matsumura, Peale, \& Rasio}]{Matsumura_2010}
Matsumura, S., Peale, S.~J., \& Rasio, F.~A. 2010, The Astrophysical Journal,
  725, 1995

\bibitem[{{Meibom} {et~al.}(2015){Meibom}, {Barnes}, {Platais}, {Gilliland},
  {Latham}, \& {Mathieu}}]{meibom15}
{Meibom}, S., {Barnes}, S.~A., {Platais}, I., {et~al.} 2015, \nat, 517, 589

\bibitem[{{Meibom} \& {Mathieu}(2005)}]{meibom}
{Meibom}, S. \& {Mathieu}, R.~D. 2005, \apj, 620, 970

\bibitem[{{Millholland} \& {Laughlin}(2018)}]{millholland}
{Millholland}, S. \& {Laughlin}, G. 2018, \apjl, 869, L15

\bibitem[{{Miralda-Escud{\'e}}(2002)}]{miralda}
{Miralda-Escud{\'e}}, J. 2002, \apj, 564, 1019

\bibitem[{{Morton}(2015)}]{morton}
{Morton}, T.~D. 2015, {isochrones: Stellar model grid package}, Astrophysics
  Source Code Library, record ascl:1503.010

\bibitem[{{Nagasawa} {et~al.}(2008){Nagasawa}, {Ida}, \& {Bessho}}]{nagasawa}
{Nagasawa}, M., {Ida}, S., \& {Bessho}, T. 2008, \apj, 678, 498

\bibitem[{{Nascimbeni} {et~al.}(2013){Nascimbeni}, {Cunial}, {Murabito},
  {Sada}, {Aparicio}, {Piotto}, {Bedin}, {Milone}, {Rosenberg}, {Zurlo},
  {Borsato}, {Damasso}, {Granata}, \& {Malavolta}}]{nascimbeni2013}
{Nascimbeni}, V., {Cunial}, A., {Murabito}, S., {et~al.} 2013, \aap, 549, A30

\bibitem[{{Nascimbeni} {et~al.}(2011){Nascimbeni}, {Piotto}, {Bedin}, \&
  {Damasso}}]{nascim}
{Nascimbeni}, V., {Piotto}, G., {Bedin}, L.~R., \& {Damasso}, M. 2011, \aap,
  527, A85

\bibitem[{{Nascimbeni} {et~al.}(2022){Nascimbeni}, {Piotto}, {B{\"o}rner},
  {Montalto}, {Marrese}, {Cabrera}, {Marinoni}, {Aerts}, {Altavilla},
  {Benatti}, {Claudi}, {Deleuil}, {Desidera}, {Fabrizio}, {Gizon}, {Goupil},
  {Granata}, {Heras}, {Magrin}, {Malavolta}, {Mas-Hesse}, {Ortolani}, {Pagano},
  {Pollacco}, {Prisinzano}, {Ragazzoni}, {Ramsay}, {Rauer}, \&
  {Udry}}]{nascimbeni_2022}
{Nascimbeni}, V., {Piotto}, G., {B{\"o}rner}, A., {et~al.} 2022, \aap, 658, A31

\bibitem[{{Nelson} {et~al.}(2000){Nelson}, {Papaloizou}, {Masset}, \&
  {Kley}}]{nelson}
{Nelson}, R.~P., {Papaloizou}, J. C.~B., {Masset}, F., \& {Kley}, W. 2000,
  \mnras, 318, 18

\bibitem[{{Nichols} {et~al.}(2015){Nichols}, {Wynn}, {Goad}, {Alexander},
  {Casewell}, {Cowley}, {Burleigh}, {Clarke}, \& {Bisikalo}}]{nichols}
{Nichols}, J.~D., {Wynn}, G.~A., {Goad}, M., {et~al.} 2015, \apj, 803, 9

\bibitem[{{Ogilvie} \& {Lin}(2007)}]{olgivelin}
{Ogilvie}, G.~I. \& {Lin}, D.~N.~C. 2007, \apj, 661, 1180

\bibitem[{{{\"O}zt{\"u}rk} \& {Erdem}(2019)}]{ozturk}
{{\"O}zt{\"u}rk}, O. \& {Erdem}, A. 2019, \mnras, 486, 2290

\bibitem[{{Parmentier} {et~al.}(2018){Parmentier}, {Line}, {Bean}, {Mansfield},
  {Kreidberg}, {Lupu}, {Visscher}, {D{\'e}sert}, {Fortney}, {Deleuil},
  {Arcangeli}, {Showman}, \& {Marley}}]{parementier}
{Parmentier}, V., {Line}, M.~R., {Bean}, J.~L., {et~al.} 2018, \aap, 617, A110

\bibitem[{Patra {et~al.}(2020)Patra, Winn, Holman, Gillon, Burdanov, Jehin,
  Delrez, Pozuelos, Barkaoui, Benkhaldoun, Narita, Fukui, Kusakabe, Kawauchi,
  Terada, Bouma, Weinberg, \& Broome}]{patra}
Patra, K.~C., Winn, J.~N., Holman, M.~J., {et~al.} 2020, 159, 150

\bibitem[{Patra {et~al.}(2017)Patra, Winn, Holman, Yu, Deming, \&
  Dai}]{patra2017}
Patra, K.~C., Winn, J.~N., Holman, M.~J., {et~al.} 2017, 154, 4

\bibitem[{{Paxton} {et~al.}(2011){Paxton}, {Bildsten}, {Dotter}, {Herwig},
  {Lesaffre}, \& {Timmes}}]{paxton}
{Paxton}, B., {Bildsten}, L., {Dotter}, A., {et~al.} 2011, \apjs, 192, 3

\bibitem[{{Penev} {et~al.}(2018){Penev}, {Bouma}, {Winn}, \& {Hartman}}]{penev}
{Penev}, K., {Bouma}, L.~G., {Winn}, J.~N., \& {Hartman}, J.~D. 2018, \aj, 155,
  165

\bibitem[{{Poddan{\'y}} {et~al.}(2010){Poddan{\'y}}, {Br{\'a}t}, \&
  {Pejcha}}]{poddan}
{Poddan{\'y}}, S., {Br{\'a}t}, L., \& {Pejcha}, O. 2010, \na, 15, 297

\bibitem[{{Rasio} {et~al.}(1996){Rasio}, {Tout}, {Lubow}, \& {Livio}}]{rasio}
{Rasio}, F.~A., {Tout}, C.~A., {Lubow}, S.~H., \& {Livio}, M. 1996, \apj, 470,
  1187

\bibitem[{{Ros{\'a}rio} {et~al.}(2022){Ros{\'a}rio}, {Barros}, {Demangeon}, \&
  {Santos}}]{rosario}
{Ros{\'a}rio}, N.~M., {Barros}, S.~C.~C., {Demangeon}, O.~D.~S., \& {Santos},
  N.~C. 2022, \aap, 668, A114

\bibitem[{{Sada} {et~al.}(2012){Sada}, {Deming}, {Jennings}, {Jackson},
  {Hamilton}, {Fraine}, {Peterson}, {Haase}, {Bays}, {Lunsford}, \&
  {O'Gorman}}]{sada}
{Sada}, P.~V., {Deming}, D., {Jennings}, D.~E., {et~al.} 2012, \pasp, 124, 212

\bibitem[{{Sasselov}(2003)}]{sasselov}
{Sasselov}, D.~D. 2003, \apj, 596, 1327

\bibitem[{Schwarz(1978)}]{gideon}
Schwarz, G. 1978, The Annals of Statistics, 6, 461

\bibitem[{{Sestito} \& {Randich}(2005)}]{Sestitoetal2005}
{Sestito}, P. \& {Randich}, S. 2005, \aap, 442, 615

\bibitem[{{Silburt} {et~al.}(2015){Silburt}, {Gaidos}, \& {Wu}}]{silburt}
{Silburt}, A., {Gaidos}, E., \& {Wu}, Y. 2015, \apj, 799, 180

\bibitem[{Skrutskie {et~al.}(2006)Skrutskie, Cutri, Stiening, Weinberg,
  Schneider, Carpenter, Beichman, Capps, Chester, Elias, Huchra, Liebert,
  Lonsdale, Monet, Price, Seitzer, Jarrett, Kirkpatrick, Gizis, Howard, Evans,
  Fowler, Fullmer, Hurt, Light, Kopan, Marsh, McCallon, Tam, Dyk, \&
  Wheelock}]{skrutskie_2006}
Skrutskie, M.~F., Cutri, R.~M., Stiening, R., {et~al.} 2006, The Astronomical
  Journal, 131, 1163

\bibitem[{{Sneden}(1973)}]{Sneden1973}
{Sneden}, C. 1973, \aj, 184, 839

\bibitem[{{Southworth} {et~al.}(2022){Southworth}, {Barker}, {Hinse}, {Jongen},
  {Dominik}, {J{\o}rgensen}, {Longa-Pe{\~n}a}, {Sajadian}, {Snodgrass},
  {Tregloan-Reed}, {Bach-M{\o}ller}, {Bonavita}, {Bozza}, {Burgdorf}, {Figuera
  Jaimes}, {Helling}, {Hitchcock}, {Hundertmark}, {Khalouei}, {Korhonen},
  {Mancini}, {Peixinho}, {Rahvar}, {Rabus}, {Skottfelt}, \&
  {Spyratos}}]{southworth2022}
{Southworth}, J., {Barker}, A.~J., {Hinse}, T.~C., {et~al.} 2022, \mnras, 515,
  3212

\bibitem[{{Stangret} {et~al.}(2022){Stangret}, {Casasayas-Barris}, {Pall{\'e}},
  {Orell-Miquel}, {Morello}, {Luque}, {Nowak}, \& {Yan}}]{monika}
{Stangret}, M., {Casasayas-Barris}, N., {Pall{\'e}}, E., {et~al.} 2022, \aap,
  662, A101

\bibitem[{{Stevenson} {et~al.}(2014){Stevenson}, {Bean}, {Seifahrt},
  {D{\'e}sert}, {Madhusudhan}, {Bergmann}, {Kreidberg}, \&
  {Homeier}}]{stevenson}
{Stevenson}, K.~B., {Bean}, J.~L., {Seifahrt}, A., {et~al.} 2014, \aj, 147, 161

\bibitem[{{Taylor}(2005)}]{Taylor2005}
{Taylor}, M.~B. 2005, in Astronomical Society of the Pacific Conference Series,
  Vol. 347, Astronomical Data Analysis Software and Systems XIV, ed.
  P.~{Shopbell}, M.~{Britton}, \& R.~{Ebert}, 29

\bibitem[{{Taylor}(2006)}]{Taylor2006}
{Taylor}, M.~B. 2006, in Astronomical Society of the Pacific Conference Series,
  Vol. 351, Astronomical Data Analysis Software and Systems XV, ed.
  C.~{Gabriel}, C.~{Arviset}, D.~{Ponz}, \& S.~{Enrique}, 666

\bibitem[{{Turner} {et~al.}(2022){Turner}, {Flagg}, {Ridden-Harper}, \&
  {Jayawardhana}}]{turner22}
{Turner}, J.~D., {Flagg}, L., {Ridden-Harper}, A., \& {Jayawardhana}, R. 2022,
  \aj, 163, 281

\bibitem[{{Turner} {et~al.}(2021){Turner}, {Ridden-Harper}, \&
  {Jayawardhana}}]{turner}
{Turner}, J.~D., {Ridden-Harper}, A., \& {Jayawardhana}, R. 2021, \aj, 161, 72

\bibitem[{Vissapragada {et~al.}(2022)Vissapragada, Chontos, Greklek-McKeon,
  Knutson, Dai, González, Grunblatt, Huber, \& Saunders}]{Vissapragada_2022}
Vissapragada, S., Chontos, A., Greklek-McKeon, M., {et~al.} 2022, The
  Astrophysical Journal Letters, 941, L31

\bibitem[{{Weinberg} {et~al.}(2024){Weinberg}, {Davachi}, {Essick}, {Yu},
  {Arras}, \& {Belland}}]{weinberg2023}
{Weinberg}, N.~N., {Davachi}, N., {Essick}, R., {et~al.} 2024, \apj, 960, 50

\bibitem[{{Weinberg} {et~al.}(2017){Weinberg}, {Sun}, {Arras}, \&
  {Essick}}]{weinberg}
{Weinberg}, N.~N., {Sun}, M., {Arras}, P., \& {Essick}, R. 2017, \apjl, 849,
  L11

\bibitem[{{Wenger} {et~al.}(2000){Wenger}, {Ochsenbein}, {Egret}, {Dubois},
  {Bonnarel}, {Borde}, {Genova}, {Jasniewicz}, {Lalo{\"e}}, {Lesteven}, \&
  {Monier}}]{Wenger2000}
{Wenger}, M., {Ochsenbein}, F., {Egret}, D., {et~al.} 2000, \aaps, 143, 9

\bibitem[{{Winn}(2010)}]{winn_2010}
{Winn}, J.~N. 2010, in Exoplanets, ed. S.~{Seager}, 55--77

\bibitem[{{Winn} {et~al.}(2007{\natexlab{a}}){Winn}, {Holman}, {Bakos},
  {P{\'a}l}, {Johnson}, {Williams}, {Shporer}, {Mazeh}, {Fernandez}, {Latham},
  \& {Gillon}}]{winn_2007a}
{Winn}, J.~N., {Holman}, M.~J., {Bakos}, G.~{\'A}., {et~al.}
  2007{\natexlab{a}}, \aj, 134, 1707

\bibitem[{{Winn} {et~al.}(2007{\natexlab{b}}){Winn}, {Holman}, {Henry},
  {Roussanova}, {Enya}, {Yoshii}, {Shporer}, {Mazeh}, {Johnson}, {Narita}, \&
  {Suto}}]{winn_2007b}
{Winn}, J.~N., {Holman}, M.~J., {Henry}, G.~W., {et~al.} 2007{\natexlab{b}},
  \aj, 133, 1828

\bibitem[{{Wong} {et~al.}(2022){Wong}, {Shporer}, {Vissapragada},
  {Greklek-McKeon}, {Knutson}, {Winn}, \& {Benneke}}]{wong}
{Wong}, I., {Shporer}, A., {Vissapragada}, S., {et~al.} 2022, \aj, 163, 175

\bibitem[{Wong {et~al.}(2021)Wong, Shporer, Zhou, Kitzmann, Komacek, Tan,
  Tronsgaard, Buchhave, Vissapragada, Greklek-McKeon, Rodriguez, Ahlers, Quinn,
  Furlan, Howell, Bieryla, Heng, Knutson, Collins, McLeod, Berlind, Brown,
  Calkins, de~Leon, Esparza-Borges, Esquerdo, Fukui, Gan, Girardin, Gnilka,
  Ikoma, Jensen, Kielkopf, Kodama, Kurita, Lester, Lewin, Marino, Murgas,
  Narita, Pallé, Schwarz, Stassun, Tamura, Watanabe, Benneke, Ricker, Latham,
  Vanderspek, Seager, Winn, Jenkins, Caldwell, Fong, Huang, Mireles, Schlieder,
  Shiao, \& Villaseñor}]{Wong_2021}
Wong, I., Shporer, A., Zhou, G., {et~al.} 2021, The Astronomical Journal, 162,
  256

\bibitem[{{Wright} {et~al.}(2010){Wright}, {Eisenhardt}, {Mainzer}, {Ressler},
  {Cutri}, {Jarrett}, {Kirkpatrick}, {Padgett}, {McMillan}, {Skrutskie},
  {Stanford}, {Cohen}, {Walker}, {Mather}, {Leisawitz}, {Gautier}, {McLean},
  {Benford}, {Lonsdale}, {Blain}, {Mendez}, {Irace}, {Duval}, {Liu}, {Royer},
  {Heinrichsen}, {Howard}, {Shannon}, {Kendall}, {Walsh}, {Larsen}, {Cardon},
  {Schick}, {Schwalm}, {Abid}, {Fabinsky}, {Naes}, \& {Tsai}}]{wright_2010}
{Wright}, E.~L., {Eisenhardt}, P. R.~M., {Mainzer}, A.~K., {et~al.} 2010, \aj,
  140, 1868

\bibitem[{{Wu} \& {Lithwick}(2011)}]{wu}
{Wu}, Y. \& {Lithwick}, Y. 2011, \apj, 735, 109

\bibitem[{{Yang} \& {Chary}(2022)}]{yang}
{Yang}, F. \& {Chary}, R.-R. 2022, \aj, 164, 259

\bibitem[{{Yee} {et~al.}(2020){Yee}, {Winn}, {Knutson}, {Patra},
  {Vissapragada}, {Zhang}, {Holman}, {Shporer}, \& {Wright}}]{yee}
{Yee}, S.~W., {Winn}, J.~N., {Knutson}, H.~A., {et~al.} 2020, \apjl, 888, L5

\bibitem[{{Zahn}(1977)}]{zahn}
{Zahn}, J.~P. 1977, \aap, 57, 383

\bibitem[{{Zahn}(2008)}]{zahn08}
{Zahn}, J.~P. 2008, in EAS Publications Series, Vol.~29, EAS Publications
  Series, ed. M.~J. {Goupil} \& J.~P. {Zahn}, 67--90

\end{thebibliography}

\newpage

\begin{appendix}
\clearpage
\onecolumn

\section{Stacked $r$ and $R$-band light curves, with the best fit-model overplotted and binned data}

\begin{figure*}[!h]
    \centering
    \includegraphics[width=\textwidth,height=\textheight,keepaspectratio]{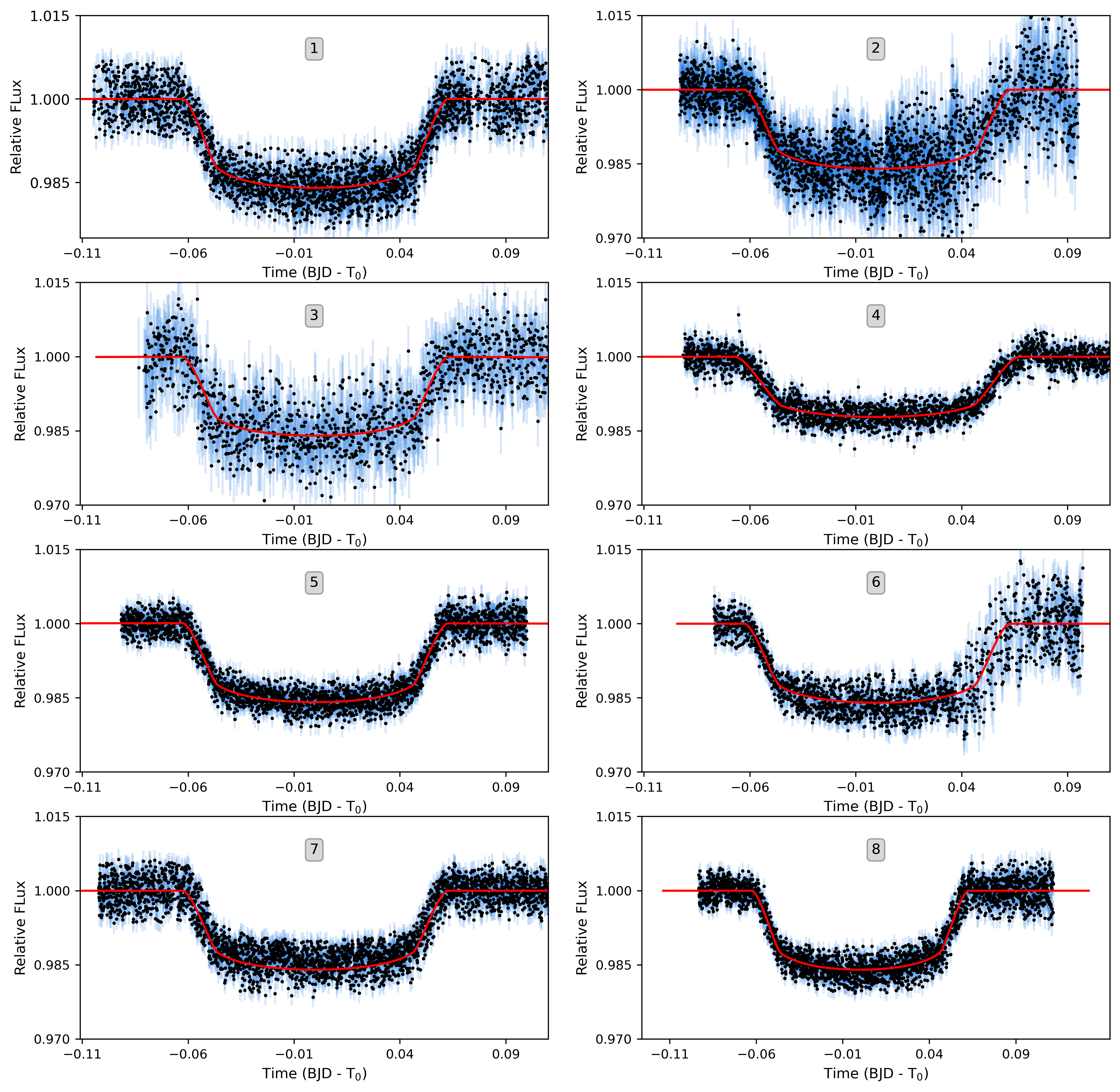}
    \caption{Individual TASTE transits light curves of WASP-12 b. The best-fitting PyORBIT model (obtained in Sect. \ref{section_fitting}) is shown as a red line. The ID for each transit is shown in the center of each panel. }
    \label{figure:firstlc}
\end{figure*}
\begin{figure*}[!h]
    \centering
    \includegraphics[width=\textwidth,height=\textheight,keepaspectratio]{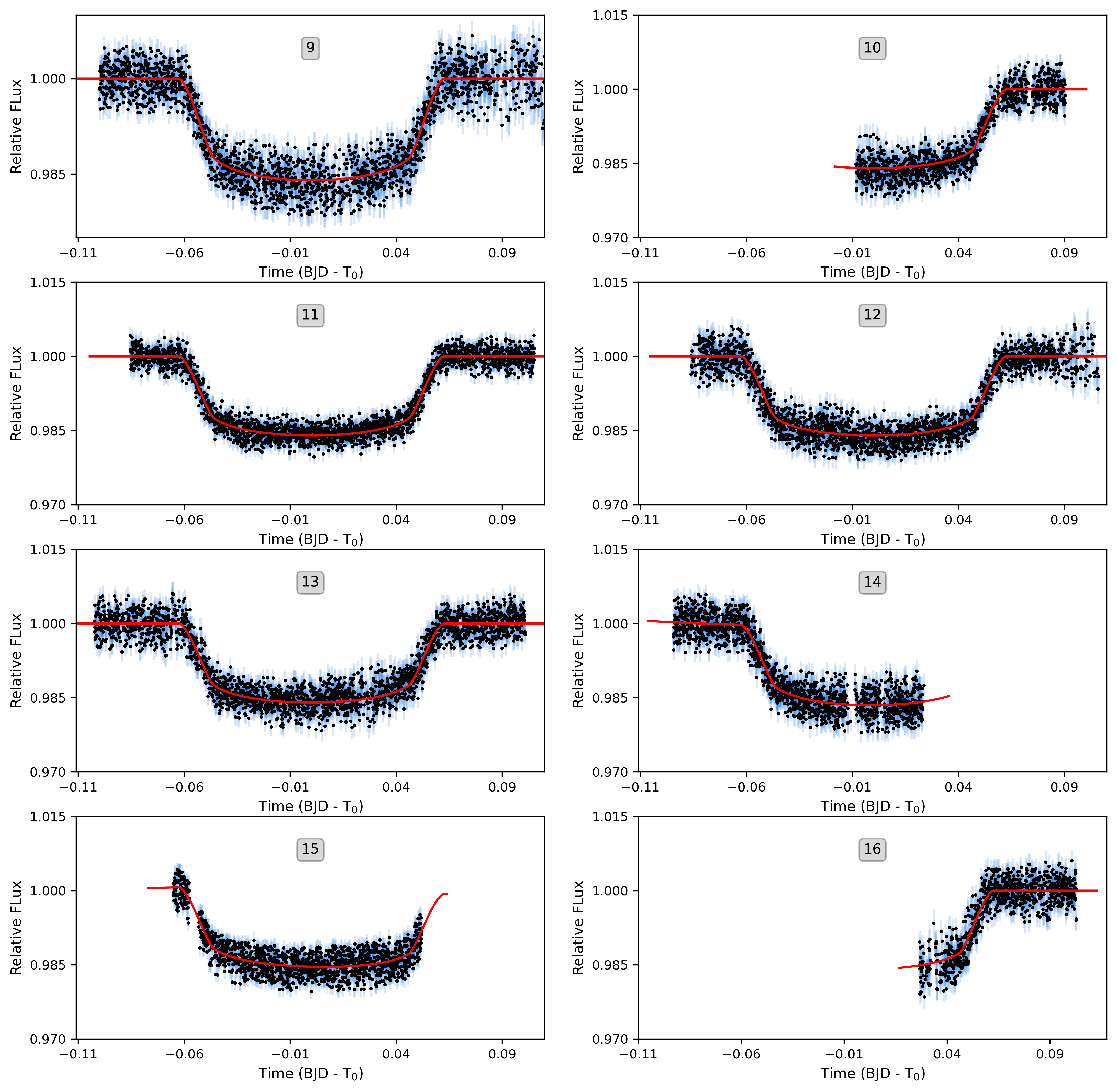}
    \caption{Same as Figure \ref{figure:firstlc} but for observations 9 through 16.}
    \label{figure:secondlc}
\end{figure*}
\begin{figure*}[!h]
    \centering
    \includegraphics[width=\textwidth,height=\textheight,keepaspectratio]{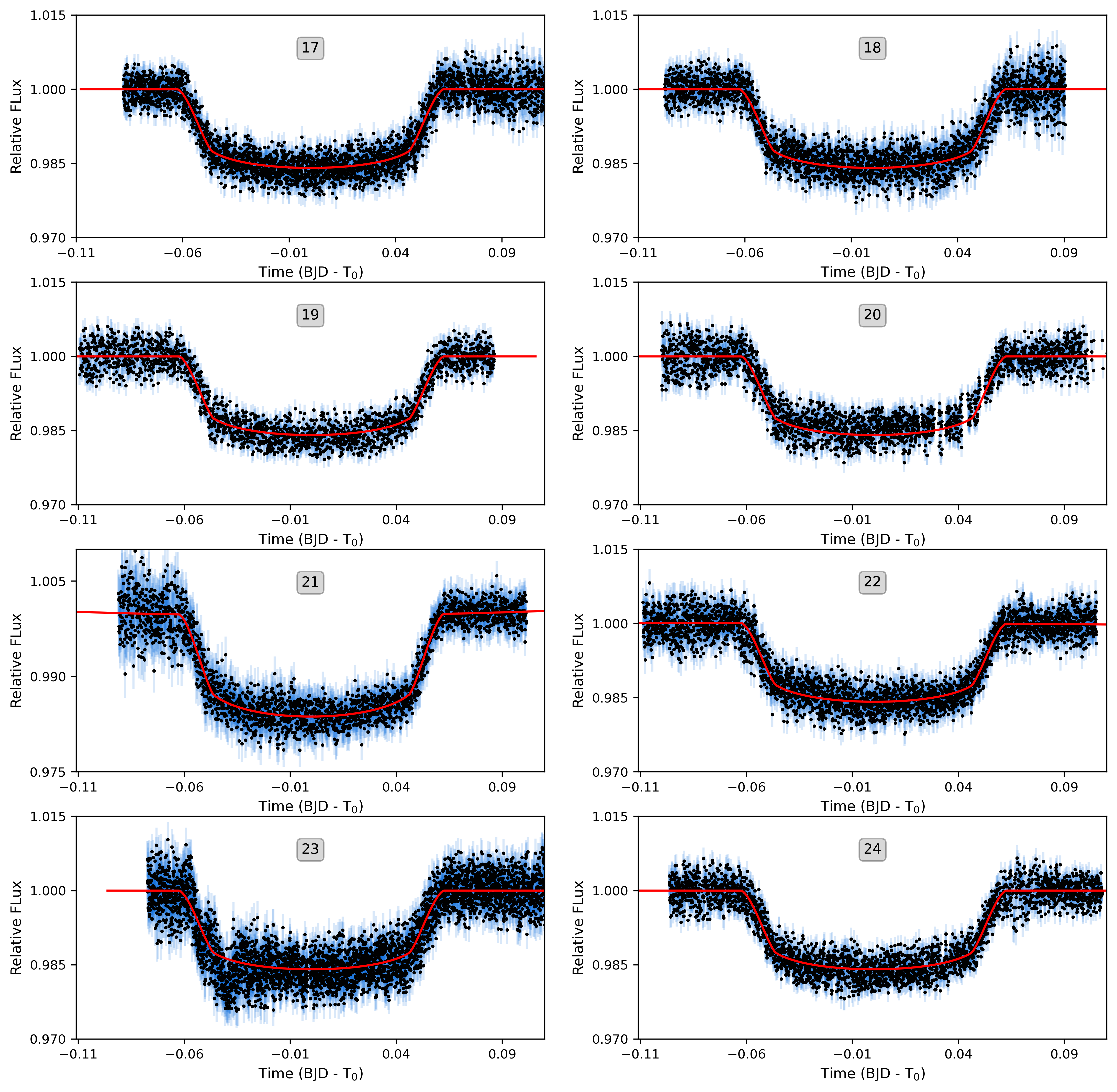}
    \caption{Same as Figure \ref{figure:firstlc} but for observations 17 through 24.}
    \label{figure:thirdlc}
\end{figure*}

\begin{figure*}[!h]
    \centering
    \includegraphics[width=\textwidth,height=\textheight,keepaspectratio]{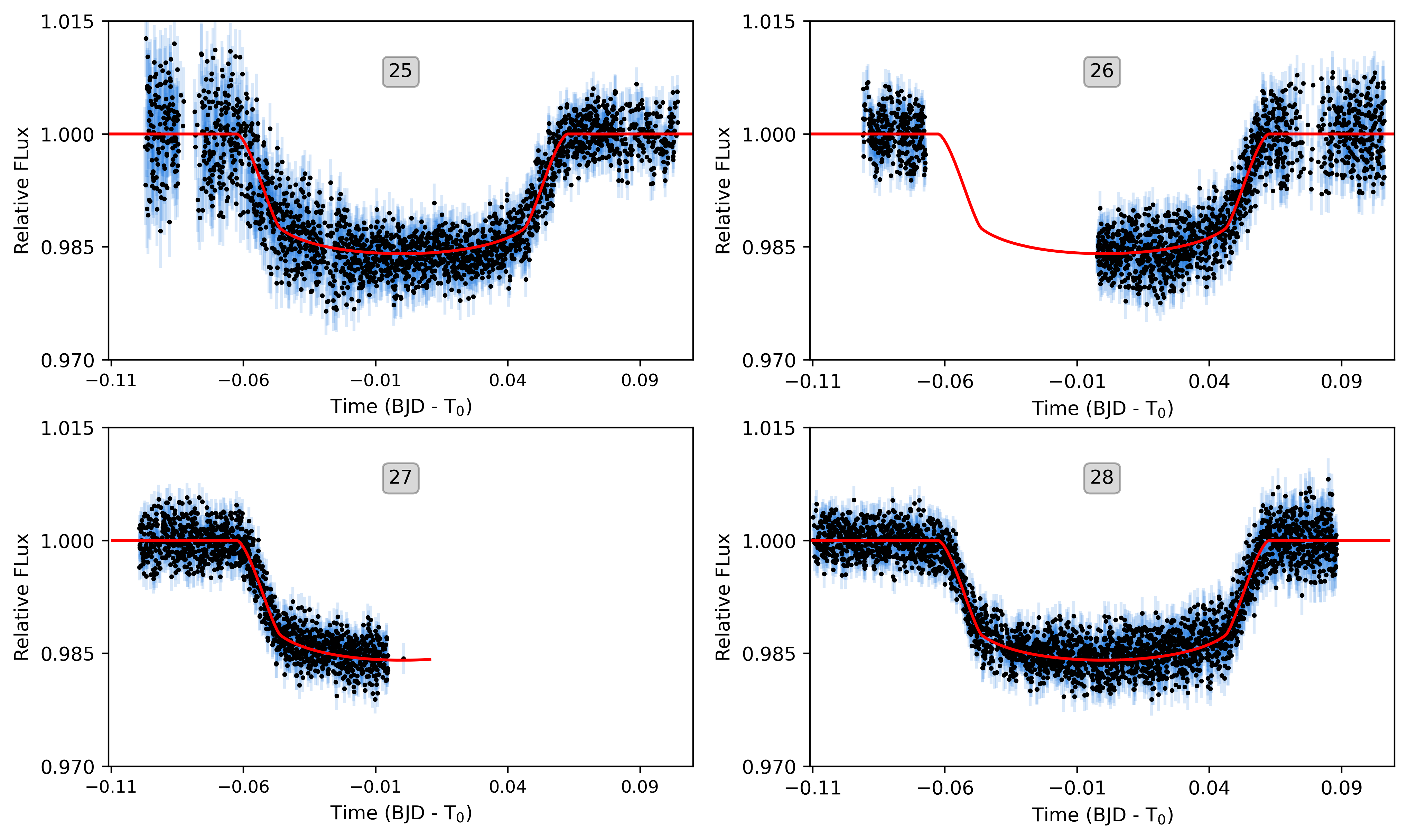}
    \caption{Same as Figure \ref{figure:firstlc} but for observations 25 through 28. }
    \label{figure:lastlc}
\end{figure*}

\section{H-R diagram}

\begin{figure*}[!h]
    \centering
    \includegraphics[width=\textwidth,height=\textheight,keepaspectratio]{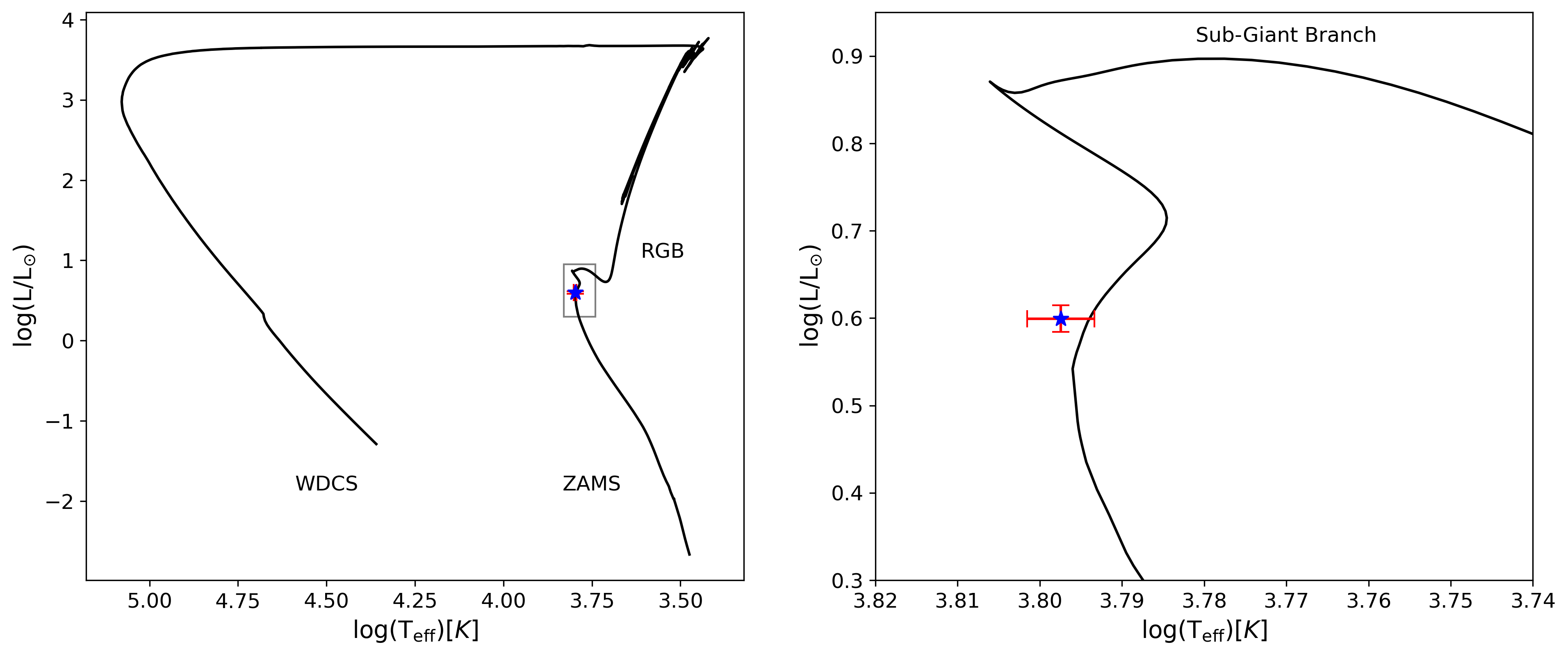}
    \caption{H-R diagram displaying the effective temperature $T_{eff}$ versus luminosity $L/L_{\odot}$, the isochrone from the  MESA Isochrone \& Stellar Tracks (MIST) stellar evolution models is shown as a solid black line. The location of WASP-12 is shown with a filled blue star. The gray box marks the zoomed-in region shown in the right panel. Right: zoomed in view of the diagram.}
    \label{figure:H-R}
\end{figure*}

\end{appendix}
\end{document}